\documentclass[12pt,onecolumn,doublespace]{autart}
\usepackage{graphics} 
\usepackage{epsfig} 
\usepackage{mathptmx} 
\usepackage{times} 
\usepackage{amsmath} 
\usepackage{amssymb} 
\usepackage{enumerate}
\usepackage{mathrsfs}
\usepackage{multirow}
\usepackage{subfigure}
\usepackage{latexsym}
\usepackage{bm}
\usepackage{dsfont}
\usepackage{color}
\usepackage{url}
\usepackage{multirow}
\usepackage{multicol}

\newtheorem{remark}{Remark}
\newtheorem{assumption}{Assumption}
\newtheorem{lemma}{Lemma}

\newtheorem{definition}{Definition}
\newtheorem{corollary}{Corollary}
\newtheorem{theorem}{Theorem}

\graphicspath{{figures/}}

{
\begin{bmatrix}}%
{\end{bmatrix}
}

\def\build#1_#2^#3{\mathrel{\mathop{\kern0pt#1}\limits_{#2}^{#3}}}%

\makeatother

\newcommand{\MF}[1]{\textcolor{black}{#1}}
\newcommand{\GFT}[1]{\textcolor{black}{#1}}

\begin{document}

\begin{frontmatter}
\title{Moving horizon partition-based state estimation of large-scale systems - Revised version}

\author[Milano]{Marcello Farina}
\author[Pavia]{Giancarlo Ferrari-Trecate}
\author[Milano]{Riccardo Scattolini}

\address[Milano]{Dipartimento di Elettronica e Informazione, Politecnico di Milano, Via Ponzio 34/5, 20133, Milan, Italy, {\tt\small \{farina,riccardo.scattolini\}@elet.polimi.it}}
\address[Pavia]{Dipartimento di Informatica e Sistemistica, Universita' degli Studi di Pavia, Via Ferrata 1, 27100 Pavia, Italy,
        {\tt\small giancarlo.ferrari@unipv.it}}

\begin{abstract}
This report presents three Moving Horizon Estimation (MHE) methods for discrete-time partitioned linear systems, i.e. systems decomposed into coupled subsystems with non-overlapping states. The MHE approach is used due to its capability of exploiting physical constraints on states in the estimation process. In the proposed algorithms, each subsystem solves reduced-order MHE problems to estimate its own state and different estimators have different computational complexity, accuracy and transmission requirements among subsystems. In all cases, conditions for the convergence of the estimation error to zero are analyzed.
\end{abstract}

\end{frontmatter}

\maketitle
\section{Introduction}
Decentralized state-estimation algorithms for large-scale systems decomposed into physically coupled subsystems is of paramount importance in many engineering control problems, such as power networks~\cite{Siljac78}, transport networks~\cite{s1978} and process control~\cite{Vadigepalli2003}. For this reason, many studies focused on the design of decentralized Kalman filters and the different solutions proposed can be classified according to the model used by each subsystem for state-estimation purposes and to the topology of the communication network among subsystems. Early works, e.g. \cite{Hashemipour1988}, \cite{Rao1991} aimed at reducing the computational complexity of centralized Kalman filtering by parallelizing computations. However, they require all-to-all communication and assume each subsystem has full knowledge of the whole dynamics. In \cite{Mutambarac1998} the focus is on the use of reduced-order and decoupled models for each subsystem. The proposed solutions, beside neglecting coupling, exploit communication networks that are almost fully connected. Subsystems with overlapping states have been considered in \cite{Khan2008}, \cite{Vadigepalli2003},~\cite{Stankovic09_TAC} and~\cite{Stankovic09_Automatica}. While the estimation schemes in \cite{Vadigepalli2003} require all-to-all communication, in \cite{Khan2008},~\cite{Stankovic09_TAC} and~\cite{Stankovic09_Automatica} the topology of the network is defined by dependencies among the states of subsystems resulting in a fully decentralized scheme. The extreme case results in distributed state-estimation where the whole process is observed by a sensor network and each sensor measures just some of the system outputs. In this case, as shown in \cite{Rantzer06,Saber07CDC,Carli07,Tomlin08,FFS09} convergence of the estimates to a common value can be achieved through consensus algorithms under weak assumptions on the topology of the communication network.

One drawback of (centralized or decentralized) Kalman filtering is that known physical constraints on state variables are not exploited in the estimation process. This can lead to suboptimal estimates or instability of the error dynamics \cite{RaoPhD00}.
 In order to overcome these issues, Moving Horizon Estimation (MHE) has been proposed for discrete-time linear \cite{Rao01,ABB03},
 nonlinear \cite{Rao03,ABB08,ABPZ99} and hybrid systems \cite{Ferrari-Trecate2002}. MHE amounts to solve at each time instant an
 optimization problem whose complexity scales with the number of states, inputs and the estimation horizon. \MF{While the algorithms proposed in~\cite{ABB03} and~\cite{ABB08} are developed for unconstrained systems affected by bounded disturbances, the methods proposed in~\cite{Rao01} and~\cite{Rao03} are capable to cope with state constraints, at the price of greater computational complexity.}

In this report we propose three Partition-based MHE algorithms (PMHE), namely PMHE1, PMHE2 and PMHE3, for linear constrained systems that are decomposed into interconnected subsystems without overlapping states. In all cases each subsystem solves a reduced-order MHE problem in order to estimate its own states. The proposed solutions have different features in terms of communication requirements among subsystems, accuracy and computational complexity.
While PMHE1 and PMHE2 provide a decentralization of the MHE scheme proposed in \cite{Rao01}, PMHE3 is inspired to the MHE strategy for unconstrained systems described in \cite{ABB03}. Decentralization is achieved through suitable approximations of covariance matrices and results in suboptimal estimation algorithms, compared to centralized MHE. Nevertheless we provide conditions guaranteeing convergence of the PMHE schemes.\\

An earlier version of the results reported in this report has been already published in \cite{FAFTSC}. However, in \cite{FAFTSC}, and specifically in the verification of the recursive feasibility property of the underlying optimization problems to be solved at any time instant, an error occurred and propagated. This bug has been fixed here by properly reformulating the three algorithms and their convergence properties have been properly established. These modifications have lead to slightly different formulations of the optimization problems to be recursively solved, the most significant of which is that only state constraints can now be imposed in PMHE1 and PMHE2, while PMHE3 turns out to be unconstrained.

\medskip
\textbf{Notation.} $I_n$ and $0$ denote the
$n\times n$ identity matrix and the matrix of zero elements whose dimensions will be clear from the context, respectively. The notation $\|z\|^2_S$ stands for $z^TSz$, where $S$ is a symmetric positive-semidefinite matrix. We use the short-hand $\mathbf{v}=(v_1,\dots,v_s)$ to denote a column vector with $s$ (not necessarily scalar) components. Finally, we define the operators
\begin{subequations}
    \label{eq:opt_eqs}
    \begin{align}
        \mathcal{R}(P,C,R)&=P-PC^T(CPC^T+R)^{-1}CP=(P^{-1}+C^TR^{-1}C)^{-1}\\
        \mathcal{R}^+(P,A,Q)&=APA^T+Q\\
        \mathcal{K}(P,C,R)&=PC^T(CPC^T+R)^{-1}=(P^{-1}+C^TR^{-1}C)^{-1}C^T R^{-1}
    \end{align}
\end{subequations}
where $\mathcal{R}$ stands for ''Riccati'' and $\mathcal{K}$ provides the Kalman gain.

\section{Partitioned systems}
\label{sec:pb-form}
Consider the discrete-time linear system
\begin{equation} \mathbf{x}_{t+1}=\mathbf{A}\,\mathbf{x}_t+\mathbf{w}_t, \label{eq:state_eq_0}\end{equation}
where $\mathbf{x}_t\in \mathbb{R}^{n}$ is the state vector, while
$\mathbf{w}_t$ represents a disturbance with variance $\mathbf{Q}^o> 0$. The initial condition $\mathbf{x}_0$ is a random
variable with mean $\mathbf{m}_{x_0}$ and covariance matrix $\bm{\Pi}_0>0$.
Measurements on the state vector are performed according to the sensing model
\begin{equation}\mathbf{y}_{t}=\mathbf{C}\,\mathbf{x}_t+\mathbf{v}_t\label{eq:output_equation_collective}
\end{equation}
where $\mathbf{v}_t\in\mathbb{R}^{p}$ is a white noise with variance $\mathbf{R}^o>0$.

Let system \eqref{eq:state_eq_0} be partitioned in $M$ low order interconnected submodels with non-overlapping states i.e., where a generic submodel has $x^{[i]}_t\in\mathbb{R}^{n_i}$ as state vector, and $\mathbf{x}_t=(x_t^{[1]},\dots,x_t^{[M]})$. Accordingly, the state transition matrices $A^{[1]}\in\mathbb{R}^{n_1\times n_1}$, $\dots$, $A^{[M]}\in\mathbb{R}^{n_M\times n_M}$ of the $M$ subsystems are diagonal blocks of $\mathbf{A}$, whereas the non-diagonal blocks of $\mathbf{A}$ define the coupling terms between subsystems.
It results that the $i$-th subsystem obeys to the linear dynamics
\begin{equation} x^{[i]}_{t+1}=A^{[i]}\,x^{[i]}_t+u^{[i],x}_{t}+w^{[i]}_t, \label{eq:state_eq_sub0}\end{equation}
where $x^{[i]}_t$ is the state vector, $u^{[i],x}_{t}$ collects the effect of state variables of other subsystems (and will be specified later on), and the term
$w^{[i]}_t$ is a disturbance with variance $Q^{[i]}$. In the decomposition we also assume that state of the $i$-th subsystem verifies the bound $x^{[i]}_t\in\mathbb{X}_i$, where $\mathbb{X}_i$ is a convex set. Correspondingly, constraints for model~\eqref{eq:state_eq_0} are $\mathbf{x}_t\in\mathbb{X}=\Pi_{i=1}^M\mathbb{X}_i$. When $\mathbb{X}= \mathbb{R}^{n}$ we say that the system is unconstrained.
The initial condition ${x}^{[i]}_0$ is a random
variable with mean ${m}_{x_0}^{[i]}$ and covariance matrix $\Pi^{[i]}_0$.

According to~\eqref{eq:output_equation_collective} and to the state partition, the outputs of the subsystems are
given by
\begin{equation}y^{[i]}_{t}={C}^{[i]}\,{x}^{[i]}_t+u^{[i],y}_{t}+{v}^{[i]}_t\label{eq:output_equation_i}
\end{equation}
where $u^{[i],y}_{t}$ collects the \GFT{effect} of the state variables of other subsystems (it will be specified later on), and the term ${v}_t^{[i]}\in\mathbb{R}^{p_i}$ represents white noise with variance equal to ${R}^{[i]}$.
Notice that, in general, some outputs of the system~\eqref{eq:state_eq_0} can be considered as outputs of more than one subsystem, \emph{i.e.}, $\bar{p}=\sum_{i=1}^m p_i\geq p$. We now define $\mathbf{y}^{\sharp}_t=(y^{[1]}_{t},\dots,y^{[M]}_{t})$. Accordingly, there exists a matrix $\mathbf{H}\in\mathbb{R}^{\bar{p}\times p}$ with rank $p$, such that $\mathbf{y}^{\sharp}_t=\mathbf{H}{\mathbf{y}}_t$ \GFT{$\forall \mathbf{y_t} \in \mathbb{R}^{p}$}.
We set $\mathbf{C}^{\sharp}=\mathbf{H}\mathbf{C}$ and $\mathbf{v}^{\sharp}_t=\mathbf{H}{\mathbf{v}}_t$. \MF{Note that $Q^{[i]}> 0$, $R^{[i]}>0$ and $\Pi^{[i]}_0>0$ can be obtained from $\mathbf{Q}^o$, $\mathbf{R}^o$ and $\bm{\Pi}_0$, e.g., see~\cite{Stankovic09_Automatica}.}
From now on, we assume that the system partitioning has been carried out in such a way that the following assumption \GFT{holds}.

\begin{assumption}
The pairs $(A^{[i]},C^{[i]})$ are observable, for $i=1,..., M$.
\label{ass:local-observability}
\end{assumption}
Notice that, neither Assumption~\ref{ass:local-observability} implies that the pair $(\mathbf{A},\mathbf{C})$ is observable, nor observability of~\eqref{eq:state_eq_0}-\eqref{eq:output_equation_collective} implies Assumption~\ref{ass:local-observability}.

We define $n^o_i$ as the observability index of the pair $(A^{[i]},C^{[i]})$ and $\bar{n}^o=\max_{i=1,\dots,M}n^o_i$\footnote{According to~\cite{Kailath} and in view of Assumption~\ref{ass:local-observability}, ${n}^o_i$ is defined as the minimum value of $N$ such that the matrix $\begin{bmatrix}(C^{[i]})^T&\dots&(C^{[i]}(A^{[i]})^{N-1})^T\end{bmatrix}^T$ has full column rank $n_i$.}. We introduce the matrices $\mathbf{A}^*=$diag$(A^{[1]},\dots,A^{[M]})$, $\tilde{\mathbf{A}}=\mathbf{A}-\mathbf{A}^*$, where $\tilde{\mathbf{A}}$ has the structure
$\tilde{\mathbf{A}}=\begin{bmatrix}(\tilde{{A}}^{[1]})^T&\dots&(\tilde{{A}}^{[M]})^T\end{bmatrix}^T,$
and $\tilde{{A}}^{[i]}\in\mathbb{R}^{n_i\times n}$. Furthermore $\mathbf{C}^*=$diag$(C^{[1]},\dots,C^{[M]})$, $\tilde{\mathbf{C}}=\mathbf{C}^{\sharp}-\mathbf{C}^*$, where $\tilde{\mathbf{C}}$ has the structure
$\tilde{\mathbf{C}}=\begin{bmatrix}(\tilde{{C}}^{[1]})^T&\dots&(\tilde{{C}}^{[M]})^T\end{bmatrix}^T,$
and $\tilde{{C}}^{[i]}\in\mathbb{R}^{p_i\times n}$.
Correspondingly, the inputs $u^{[i],x}_{t}$ and $u^{[i],y}_{t}$ in~\eqref{eq:state_eq_sub0} and~\eqref{eq:output_equation_i} are $u^{[i],x}_{t}=\tilde{A}^{[i]}\,\mathbf{x}_t$ and $u^{[i],y}_{t}=\tilde{C}^{[i]}\,\mathbf{x}_t$.
We say that a system partition is trivial if
$M=1$.
\section{Three moving horizon partition-based algorithms}
\label{sec:PHME}
Our aim is to design, for each subsystem, an
algorithm for computing a reliable estimate $\hat{x}^{[i]}$ of $x^{[i]}$
based on the measurements ${y}^{[i]}$ and on the estimates of the crosstalk terms $u^{[i],x}$ and $u^{[i],y}$ provided by the estimators associated to the other subsystems. To this end, we propose three solutions, named PMHE1, PMHE2 and PMHE3.
\subsection{Models for estimation and transmission of information}
\label{sec:Models}
We denote with $\hat{x}_{t_1/t_2}^{[i]}$ the estimate of ${x}_{t_1}^{[i]}$ performed at time $t_2$ by subsystem $i$. Its error covariance matrix is denoted with ${\Pi}_{t_1/t_2}^{[i]}$ and we define $\hat{\mathbf{x}}_{t_1/t_2}=(\hat{{x}}_{t_1/t_2}^{[1]},\dots,\hat{{x}}_{t_1/t_2}^{[M]})$. We approximate
\begin{equation}
\text{Var}(\mathbf{x}_{t_1}-\hat{\mathbf{x}}_{t_1/t_2})\simeq\bm{\Pi}_{t_1/t_2}=\text{diag}({\Pi}_{t_1/t_2}^{[1]},\dots,{\Pi}_{t_1/t_2}^{[M]})
\label{eq:diagonal_Pi_approx}
\end{equation}
that corresponds to assume that the errors of different subsystems are uncorrelated. This approximation will allow decentralization of the centralized MHE problem. At time $t$ the estimation model is, for $k=t-N,\dots,t-1$
\begin{subequations}
\begin{align}\hat{x}^{[i]}_{k+1}=A^{[i]}\,\hat{x}^{[i]}_k+\tilde{A}^{[i]}\,\tilde{\mathbf{x}}_{k/t-1}+\hat{w}^{[i]}_{k}\label{eq:modelS1_x}\\
y^{[i]}_{k}={C}^{[i]}\,\hat{x}^{[i]}_k+\tilde{C}^{[i]}\,\tilde{\mathbf{x}}_{k/t-1}+\hat{v}^{[i]}_k\label{eq:modelS1_y}
\end{align}
\label{eq:modelS1}
\end{subequations}
and defines constraints of the PMHE estimation problem specified in the next section. In~\eqref{eq:modelS1}, $\tilde{\mathbf{x}}_{k/t-1}\in\mathbb{R}^{n}$ denotes estimates of the subsystem states available at time $t$, that will be defined later on and that can differ from $\hat{\mathbf{x}}_{k/t-1}$. The noise terms $\hat{w}^{[i]}_{k}$ and $\hat{v}^{[i]}_k$ in~\eqref{eq:modelS1} encompass both the noise appearing in the equations~\eqref{eq:state_eq_sub0},~\eqref{eq:output_equation_i} and the estimation error of the variables $u^{[i],x}_{t}$ and $u^{[i],y}_{t}$. In fact, \eqref{eq:state_eq_sub0} and \eqref{eq:output_equation_i} are equivalent to
\begin{subequations}
\begin{align}{x}^{[i]}_{k+1}=A^{[i]}\,{x}^{[i]}_k+\tilde{A}^{[i]}\,\tilde{\mathbf{x}}_{k/t-1}+
\tilde{A}^{[i]}(\mathbf{x}_{k}-\tilde{\mathbf{x}}_{k/t-1})+{w}^{[i]}_{k}\label{eq:system_i_01_x}\\
y^{[i]}_{k}={C}^{[i]}\,{x}^{[i]}_k+\tilde{C}^{[i]}\,\tilde{\mathbf{x}}_{k/t-1}+
\tilde{C}^{[i]}(\mathbf{x}_{k}-\tilde{\mathbf{x}}_{k/t-1})+{v}^{[i]}_k\label{eq:system_i_01_y}
\end{align}
\label{eq:system_i_01}
\end{subequations}
giving rise to the collective model
\begin{subequations}
\begin{align}\mathbf{x}_{k+1}=\mathbf{A}^{*}\,\mathbf{x}_k+\tilde{\mathbf{A}}\,\tilde{\mathbf{x}}_{k/t-1}+
\tilde{\mathbf{A}}(\mathbf{x}_{k}-\tilde{\mathbf{x}}_{k/t-1})+\mathbf{w}_{k}\label{eq:system_tot_02_x}\\
\mathbf{y}^{\sharp}_{k}=\mathbf{C}^*\,\mathbf{x}_k+\tilde{\mathbf{C}}\,\tilde{\mathbf{x}}_{k/t-1}+
\tilde{\mathbf{C}}(\mathbf{x}_{k}-\tilde{\mathbf{x}}_{k/t-1})+\mathbf{v}^{\sharp}_k\label{eq:system_tot_02_y}
\end{align}
\label{eq:system_tot_02}
\end{subequations}
According to \eqref{eq:system_tot_02}, the variance of $\hat{\mathbf{w}}_k=(\hat{w}^{[1]}_{k},\dots,\hat{w}^{[M]}_{k})$ and $\hat{\mathbf{v}}_k=(\hat{v}^{[1]}_{k},\dots,\hat{v}^{[M]}_{k})$ is given by
\begin{subequations}
\begin{align}Var(\hat{\mathbf{w}}_k)&=~Var(\mathbf{w}_k)+\tilde{\mathbf{A}}Var(\mathbf{x}_{k}-\tilde{\mathbf{x}}_{k/t-1})(\tilde{\mathbf{A}})^T\label{eq:Q_PMHE_both_collective}\\ Var(\hat{\mathbf{v}}_k)&=~Var(\mathbf{v}^{\sharp}_k)+\tilde{\mathbf{C}} Var(\mathbf{x}_{k}-\tilde{\mathbf{x}}_{k/t-1})(\tilde{\mathbf{C}})^T\label{eq:R_PMHE_both_collective}
\end{align}
\label{eq:Q_and_R_PMHE_both_collective}
\end{subequations}
where a block-diagonal approximation of $Var(\mathbf{w}_k)$ and $Var(\mathbf{v}^{\sharp}_k)$ is considered, i.e., $Var(\mathbf{w}_k)\simeq \mathbf{Q}=$diag$(Q^{[1]},\dots,Q^{[M]})$ and $Var(\mathbf{v}^{\sharp}_k)\simeq \mathbf{R}^{\sharp}=$diag$(R^{[1]},\dots,R^{[M]})$. Next we introduce a model for $\tilde{\mathbf{x}}_{k/t-1}$ and approximations to $Var(\hat{\mathbf{w}}_k)$ and $Var(\hat{\mathbf{v}}_k)$, that are related to different communication protocols: the first one will be used in PMHE1, while the second one will be used in PMHE2 and PMHE3.

\textbf{Model~1}: the system partition induces an interconnected network of subsystems, which can be described by a directed graph
$\mathscr{G}=(\mathcal{V},\mathcal{E})$, where the nodes in $\mathcal{V}$ are the subsystems and the edge $(j,i)$ in the set $\mathcal{E}\subseteq \mathcal{V}\times \mathcal{V}$ models that the $j$-th subsystem influences the dynamics of the $i$-th subsystem. Therefore we assume a neighbor-to-neighbor communication i.e., at time $t$, if $(j,i)\in\mathcal{E}$ then only $\hat{x}^{[j]}_{k/t-1}$ and $\Pi^{[j]}_{k/t-1}$ for $k=t-N,\dots,t-1$ can be transmitted to subsystem $i$. We set $\tilde{\mathbf{x}}_{k/t-1}=\hat{\mathbf{x}}_{k/t-1}$ and, according to \eqref{eq:Q_and_R_PMHE_both_collective} and~\eqref{eq:diagonal_Pi_approx}, the variances of $\hat{\mathbf{w}}_k$ and $\hat{\mathbf{v}}_k$ are approximated as
\MF{\begin{subequations}
\begin{align}Var(\hat{\mathbf{w}}_k)&\simeq~\mathbf{Q}+\tilde{\mathbf{A}}\bm{\Pi}_{k/t-1}(\tilde{\mathbf{A}})^T\label{eq:Q_PMHE1_collective}\\ Var(\hat{\mathbf{v}}_k)&\simeq~\mathbf{R}^{\sharp}+\tilde{\mathbf{C}} \bm{\Pi}_{k/t-1}(\tilde{\mathbf{C}})^T\label{eq:R_PMHE1_collective}
\end{align}
\label{eq:Q_and_R_PMHE1_collective}
\end{subequations}}

One can show that lower and upper bounds to~\eqref{eq:Q_PMHE1_collective} and~\eqref{eq:R_PMHE1_collective} are given by
\MF{\begin{subequations}
\label{eq:Q_and_R_PMHE1_collective_upperbound}
\begin{align}\mathbf{Q}\leq Var(\hat{\mathbf{w}}_k)&\leq~\text{diag}(\bar{Q}^{[1]}_k,\dots,\bar{Q}^{[M]}_{k})\\
\mathbf{R}^{\sharp}\leq Var(\hat{\mathbf{v}}_k)&\leq~\text{diag}(\bar{R}^{[1]}_{k},\dots,\bar{R}^{[M]}_{k})
\end{align}
\end{subequations}}
where
\begin{subequations}
\label{eq:QandR_PMHE1}
\begin{align}\bar{Q}^{[i]}_{k}&=~Q^{[i]}+\,(\tilde{A}^{[i]})\bm{\Delta}_{x,\nu}\bm{\Pi}_{k/t-1}(\tilde{A}^{[i]})^T\label{eq:QandR_PMHE1_Q}\\
\bar{R}^{[i]}_{k}&=~R^{[i]}+(\tilde{C}^{[i]}) \bm{\Delta}_{y,\nu}\bm{\Pi}_{k/t-1}(\tilde{C}^{[i]})^T\label{eq:QandR_PMHE1_R}
\end{align}
\end{subequations}
In \eqref{eq:QandR_PMHE1}, $\bm{\Delta}_{x,\nu}=$diag$(\nu_x^{[1]} I_{n_1},\dots,\nu_x^{[M]} I_{n_M})$, $\bm{\Delta}_{y,\nu}=$diag$(\nu_y^{[1]} I_{n_1},\dots,\nu_y^{[M]} I_{n_M})$ where $\nu^{[i]}_x$ and $\nu^{[i]}_y$, $i=1,\dots,M$, are defined as the cardinalities of $\mathcal{V}^{[i],\flat}_{x}=\{j\neq i\in\mathcal{V}:A_{ji}\neq 0\}$ and $\mathcal{V}^{[i],\flat}_{y}=\{j\neq i\in\mathcal{V}:C_{ji}\neq 0\}$, respectively.
Note that, in~\eqref{eq:modelS1} and~\eqref{eq:QandR_PMHE1}, the terms $\tilde{A}^{[i]}\,\hat{\mathbf{x}}_{k/t-1}$, $\tilde{C}^{[i]}\,\hat{\mathbf{x}}_{k/t-1}$, $\bar{Q}^{[i]}_{k}$ and $\bar{R}^{[i]}_{k}$ depend only upon the quantities transmitted by the neighboring subsystems $j\in\mathcal{V}^{[i]}\in\mathcal{V}=\{j\neq i:(j,i)\in \mathcal{E}\}$.

\textbf{Model~2}: we assume all-to-all communication, so that all subsystems at time $t-1$ know the vector $\hat{\mathbf{x}}_{t-N/t-1}$ and, for PMHE2, the matrix $\bm{\Pi}_{t-N/t-1}$. Accordingly, at time $t$, the $i$-th subsystems estimation model, for $k=t-N,\cdots,t-1$, is~\eqref{eq:modelS1}, where $\tilde{\mathbf{x}}_{k/t-1}=\mathbf{A}^{k-(t-N)}\,\hat{\mathbf{x}}_{t-N/t-1}$, for $k=t-N,\dots,t-1$. The noise terms $\hat{w}^{[i]}_{k}$ and $\hat{v}^{[i]}_k$ encompass now also the uncertainty characterizing the terms $\tilde{A}^{[i]}\,\tilde{\mathbf{x}}_{k/t-1}$ and $\tilde{C}^{[i]}\,\tilde{\mathbf{x}}_{k/t-1}$, respectively, and hence their covariance will be approximated by the upper and lower bounds in~\eqref{eq:Q_and_R_PMHE1_collective_upperbound}, where now matrices $\bar{Q}^{[i]}_{k}$ and $\bar{R}^{[i]}_{k}$ are given by
\begin{subequations}
\label{eq:QandR_PMHE2}
\begin{align}\bar{Q}^{[i]}_{k}&=Q^{[i]}+\tilde{A}^{[i]}\bm{\Delta}_{x,\nu}\bm{\Pi_{k/t-1}}^{ol}(\tilde{A}^{[i]})^T\\
\bar{R}^{[i]}_{k}&=R^{[i]}+\tilde{C}^{[i]}\bm{\Delta}_{y,\nu} \bm{\Pi_{k/t-1}}^{ol}(\tilde{C}^{[i]})^T
\end{align}
where, in view of the adopted transmission model, we set
\begin{align}\bm{\Pi_{t-N/t-1}}^{ol}=Var(\mathbf{x}_{t-N}-\hat{\mathbf{x}}_{t-N/t-1})=\bm{\Pi_{t-N/t-1}}\label{eq:QandR_PMHE2_PiOL0}\end{align}
and, for $k>t-N$
\begin{align}\bm{\Pi_{k/t-1}}^{ol}=&Var(\mathbf{x}_{k}-\mathbf{A}^{k-(t-N)}\hat{\mathbf{x}}_{t-N/t-1})=
Var\left(\mathbf{A}^{k-(t-N)}(\mathbf{x}_{t-N}-\hat{\mathbf{x}}_{t-N/t-1})+\sum_{i=t-N+1}^k
\mathbf{A}^{k-i}\mathbf{w}_{i-1}\right)\nonumber\\
=&\mathbf{A}^{k-(t-N)}\bm{\Pi_{t-N/t-1}}(\mathbf{A}^{k-(t-N)})^T+\sum_{i=t-N+1}^k
\mathbf{A}^{k-i}\mathbf{Q}(\mathbf{A}^{k-i})^T\label{eq:QandR_PMHE2_PiOL}\end{align}
\end{subequations}

\subsection{The PMHE estimation problems}
Given an estimation horizon $N\geq 1$, in order to perform the PMHE$r$ algorithm (with $r=1,2,3$), each node $i\in\mathcal{V}$ at time $t$ solves the constrained minimization problem MHE-$i$ defined as
\begin{equation}\Theta^{*[i]}_{r,t}=\min_{\hat{x}^{[i]}_{t-N},\hat{w}^{[i]}}J_r^{[i]}(t-N,t,\hat{x}^{[i]}_{t-N},\hat{w}^{[i]},\hat{v}^{[i]}
,\Gamma^{[i]}_{r,t-N})\label{eq:min-pb}\end{equation}
where $\hat{w}^{[i]}$ and $\hat{v}^{[i]}$ stand for $\{\hat{w}^{[i]}_k\}_{k=t-N}^{t-1}$ and $\{\hat{v}^{[i]}_k\}_{k=t-N}^{t-1}$, respectively, under the constraints
\begin{subequations}
\begin{align}
&\left\{\begin{array}{ll}\text{System~\eqref{eq:modelS1} with transmission Model~1}&\text{if } r=1\\
\text{System~\eqref{eq:modelS1} with transmission Model~2}&\text{if }r=2,3\end{array}\right.\label{eq:system_contraints}\\
&\hat{x}^{[i]}_k\in \mathbb{X}_i\text{ if }r=1,2\\
&\hat{w}^{[i]}_k=0\text{ if }r=3
\end{align}
\label{eq:update_x&v}
\end{subequations}
where $k=t-N,\dots,t$ and the local cost function $J_r^{[i]}$ is given by
\begin{equation}\begin{array}{c}J_r^{[i]}
=\sum_{k=t-N}^{t-1}L_r^{[i]}(\hat{w}^{[i]}_k,\hat{v}^{[i]}_k)+\Gamma_{r,t-N}^{[i]}(\hat{x}^{[i]}_{t-N};\hat{x}^{[i]}_{t-N/t-1})\end{array}
\label{eq:i-th cost fcn}\end{equation}
In~\eqref{eq:i-th cost fcn}, the functions $L_r^{[i]}$ and $\Gamma_{r,t-N}^{[i]}$ are the \emph{stage cost} and the \emph{initial penalty}, respectively. For PMHE1 and PMHE2 they are defined as follows
\small\begin{align}
L_r^{[i]}
=&\frac{1}{2}\|\hat{w}^{[i]}_k\|_{(Q^{[i]}_{k/t-1})^{-1}}^2+
\frac{1}{2}\|\hat{v}^{[i]}_k\|_{({R}^{[i]}_{k/t-1})^{-1}}^2 \label{eq:def-Li}\\
\Gamma_{r,t-N}^{[i]}
=&\frac{1}{2}\|\hat{x}_{t-N}^{[i]}-\hat{x}^{[i]}_{t-N/t-1}\|^2_{({\Pi_{t-N/t-1}^{[i]}})^{-1}}+\Theta^{*[i]}_{r,t-1}\label{eq:def-Gammai}\end{align}
\normalsize
with $r=1,2$. On the other hand, for PMHE3, they are defined as
\begin{align}
L_3^{[i]}
= & \frac{1}{2}\|\hat{v}^{[i]}_k\|^2 \label{eq:def-Li3}\\
\Gamma_{3,t-N}^{[i]}
= &\frac{\mu}{2}\|\hat{x}_{t-N}^{[i]}-\hat{x}^{[i]}_{t-N/t-1}\|^2+\Theta^{*[i]}_{3,t-1}\label{eq:def-Gammai3}\end{align}

In~\eqref{eq:def-Gammai} and~\eqref{eq:def-Gammai3}, $\Theta^{*[i]}_{r,t-1}$ is defined in~\eqref{eq:min-pb} and, since it is known at time $t$, it could be neglected when solving the optimization problem~\eqref{eq:min-pb}-\eqref{eq:update_x&v}. However, it is here maintained for clarity since it plays a major role in establishing the main convergence properties of the algorithms.
The positive definite symmetric matrices $\Pi_{t-N/t-1}^{[i]}$, $Q^{[i]}_{k/t-1}$ and $R^{[i]}_{k/t-1}$ are design parameters whose choice and properties are discussed next. In the sequel, $\hat{x}^{[i]}_{t-N/t}$ and
$\left\{\hat{w}^{[i]}_{k/t}\right\}_{k=t-N}^{t-1}$ are the optimizers to
\eqref{eq:min-pb} and $\hat{x}^{[i]}_{k/t}$, $k=t-N,...,t$ is the
local state sequence stemming from $\hat{x}^{[i]}_{t-N/t}$ and
$\left\{\hat{w}^{[i]}_{k/t}\right\}_{k=t-N}^{t-1}$.

\subsection{Choice of the weights $Q_{k/t-1}^{[i]}$ and $R_{k/t-1}^{[i]}$ for PMHE1 and PMHE2}
\label{sec:QeR-proc}
Ideally, one would choose $Q_{k/t-1}^{[i]}=Var(\hat{w}^{[i]}_{k})$ and $R_{k/t-1}^{[i]}=Var(\hat{v}^{[i]}_{k})$ but, as remarked in Section \ref{sec:Models}, an exact computation of the variances of $\hat{\mathbf{w}}_k$ and $\hat{\mathbf{v}}_k$ (encompassing both noise and estimation uncertainty) can not be performed in a distributed fashion. We therefore exploit upper and lower bounds given by \eqref{eq:Q_and_R_PMHE1_collective_upperbound}, \eqref{eq:QandR_PMHE1}, and \eqref{eq:QandR_PMHE2} and introduce the following alternative choices
\begin{enumerate}[I)]
\item Set $Q_{k/t-1}^{[i]}=\bar{Q}^{[i]}_k$ and $R_{k/t-1}^{[i]}=\bar{R}^{[i]}_k$ where $\bar{Q}^{[i]}_k$ and $\bar{R}^{[i]}_k$ are defined in \eqref{eq:QandR_PMHE1} and \eqref{eq:QandR_PMHE2} for PMHE1 and PMHE2, respectively. Note that, in view of the sparse structure of matrices $\tilde{\mathbf{A}}$ and $\tilde{\mathbf{C}}$ and of the block-diagonal structure of matrix $\bm{\Pi}_{k/t-1}$, for PMHE1 the computation of the weighting matrices ${Q}_{k/t-1}^{[i]}$ and ${R}_{k/t-1}^{[i]}$ is distributed, i.e., they depend from ${\Pi}^{[j]}_{k/t-1}$ only if $j$ is a neighbor of $i$.
\item Set $Q_{k/t-1}^{[i]}={Q}^{[i]}$ and $R_{k/t-1}^{[i]}={R}^{[i]}$ both for PMHE1 and for PMHE2. This is the simplest choice that can be adopted, and it is very effective for limiting the communication burden. In fact, it does not require any transmission of covariance data from a subsystem to the neighboring ones.
\end{enumerate}

\subsection{Computation of $\Pi_{t-N/t-1}^{[i]}$ for PMHE1 and PMHE2}
\label{se:update-proc}
In this section we provide the algorithms for recursively computing the weighting matrices $\Pi^{[i]}_{k/t}$, given the matrices $Q^{[i]}_{k/t-1}$ and $R^{[i]}_{k/t-1}$. As it will be discussed in Section \ref{sec:properties}, these choices of $\Pi^{[i]}_{k/t}$ are required for guaranteeing convergence of PMHE1 and PMHE2. For each estimation scheme, the update of the covariance matrices must fulfill a given matrix inequality, which in turns depends on the type of approximation adopted for $Q_{k/t-1}^{[i]}$ and $R_{k/t-1}^{[i]}$, discussed in Section \ref{sec:QeR-proc}.
\subsubsection{Update of $\Pi_{t-N/t-1}^{[i]}$ for PMHE1}
\label{sec:3.4.1}
The following fundamental linear matrix inequality must be satisfied, for each sensor:
\begin{align}
\mathcal{P}_{1,t-1}^{[i]}\leq W_{1,t-1}^{[i]}
\label{lmi:PMHE1}
\end{align}
where $\mathcal{P}_{1,t-1}^{[i]}$ and $W_{1,t-1}^{[i]}$ will be defined next depending on the adopted choices of $Q^{[i]}_{k/t-1}$ and $R^{[i]}_{k/t-1}$.\\
When covariance matrices are updated as in point (I) of Section~\ref{sec:QeR-proc} one has
\begin{align}
\mathcal{P}_{1,t-1}^{[i]}=\mathcal{P}_{1,t-1}^{[i]}(\{\Pi^{[i]}_{k/t-1}\}_{k=t-N}^{t-1}),\,
W_{1,t-1}^{[i]}=W_{1,t-1}^{[i]}(\{\Pi^{[i]}_{k/t-2}\}_{k=t-N-1}^{t-2})
\label{lmi:P_PMHE1_I}
\end{align}
and the unknowns of~\eqref{lmi:PMHE1} are matrices $\{\Pi^{[i]}_{k/t-1}\}_{k=t-N}^{t-1}$. On the other hand, when covariance matrix updates obey to point (II) of Section~\ref{sec:QeR-proc}, one has
\begin{align}
\mathcal{P}_{1,t-1}^{[i]}=\mathcal{P}_{1,t-1}^{[i]}(\Pi^{[i]}_{t-N/t-1}),\,
W_{1,t-1}^{[i]}=W_{1,t-1}^{[i]}(\Pi^{[i]}_{t-N-1/t-2})
\label{lmi:P_PMHE1_II}
\end{align}
and the only unknown is the matrix $\Pi^{[i]}_{t-N/t-1}$.\\
As it will become clearer later on, in this case~\eqref{lmi:PMHE1} plays the same role of the covariance update in Kalman filtering. Indeed, as it will be clear from the definitions of $\mathcal{P}_{1,t-1}^{[i]}$ and $W_{1,t-1}^{[i]}$, when there is no coupling between subsystems, i.e., $\tilde{\mathbf{A}}=0$ and $\tilde{\mathbf{C}}=0$, one can always fulfill~\eqref{lmi:PMHE1} (note that, in this case, the equality holds) by choosing
\begin{align}
\Pi^{[i]}_{t-N/t-1}=\mathcal{R}^+(\mathcal{R}(\Pi^{[i]}_{t-N-1/t-2},C^{[i]},R^{[i]}),A^{[i]},Q^{[i]})
\label{lmi:Kalman_UP1}
\end{align}
where operators $\mathcal{R}^+$ and $\mathcal{R}$ provide the Kalman predictor Riccati equation, as defined in the Introduction.\\\\
The matrix $W_{1,t-1}^{[i]}$ is computed, based on locally available information, as follows:
\begin{subequations}
\begin{align}W^{[i]}_{1,t}=C_{4,i}^T(\tilde{\Pi}_{t-N/t-1}^{[i]})^{-1}C_{4,i}\label{eq:Wi_PMHE1_end}\end{align}
where
$\tilde{\Pi}_{t-N/t-1}^{[i]}=$diag$({\Pi}_{t-N/t-1}^{*[i]},\mathbf{Q}_{N-1}^{[i]},\mathbf{R}_{N-1}^{[i]})$,
$\mathbf{Q}^{[i]}_{N-1}=$diag$(Q^{[i]}_{t-N+1/t-1},\dots,Q^{[i]}_{t-1/t-1})
\in\mathbb{R}^{(N-1)n_i\times (N-1)n_i}$, and\break
$\mathbf{R}^{[i]}_{N-1}=$diag$(R^{[i]}_{t-N+1/t-1},\dots,R^{[i]}_{t-1/t-1})\in\mathbb{R}^{(N-1)p_i\times(N-1)p_i}$. Furthermore
$$C_{4,i}=\begin{bmatrix}C_{1,i}\\C_{2,i}\\C_{3,i}
\end{bmatrix}$$
where
$$\begin{array}{lcl}C_{1,i}&=&\begin{bmatrix}
I_{n_i}&0&\dots&0
\end{bmatrix}\end{array},\,
\begin{array}{lcl}
C_{2,i}&=&\begin{bmatrix}
-A^{[i]}&I_{n_i}&0&\dots&0&0\\
0&-A^{[i]}&I_{n_i}&\dots&0&0\\
\vdots&\vdots&\vdots&\ddots&\vdots&\vdots\\
0&0&0&\dots&-A^{[i]}&I_{n_i}
\end{bmatrix}
\end{array},\,
\begin{array}{lcl}
C_{3,i}&=&\begin{bmatrix}
C^{[i]}&\dots&0&0\\
\vdots&\ddots&\vdots&\vdots\\
0&\dots&C^{[i]}&0
\end{bmatrix}
\end{array}
$$
$C_{1,i}\in\mathbb{R}^{n_i\times N n_i}$, $C_{2,i}\in\mathbb{R}^{(N-1) n_i\times N n_i}$, and $C_{3,i}\in\mathbb{R}^{(N-1) p_i\times N n_i}$.Finally
\begin{align}
{\Pi}_{t-N/t-1}^{*[i]}=\mathcal{R}^+(\bar{\Pi}^{[i]}_{t-N/t-1},A^{[i]},Q^{[i]}_{t-N/t-1})
\label{eq:unc_arr_cost_ind_PMHE1_4}
\end{align}
where
\begin{align}
\bar{\Pi}_{t-N/t-1}^{[i]}=\mathcal{R}(\Pi^{[i]}_{t-N/t-1},C^{[i]},R^{[i]}_{t-N/t-1})
\label{eq:unc_arr_cost_ind_PMHE1_2}
\end{align}
\label{eq:unc_arr_cost_ind_PMHE1_various}
\end{subequations}
Note that $\mathbf{Q}_{N-1}^{[i]}$ and $\mathbf{R}_{N-1}^{[i]}$ depend upon $\{\Pi^{[i]}_{k/t-2}\}_{k=t-N-1}^{t-2}$ when the covariance update in point (I) in Section~\ref{sec:QeR-proc} is used, and this justifies equation~\eqref{lmi:P_PMHE1_I} for $W_{1,t-1}^{[i]}$.\\
The matrix $\mathcal{P}_{1,t-1}^{[i]}\in\mathbb{R}^{N n_i\times N n_i}$ has different formulations depending on which approximation of the covariance matrix is chosen.
Specifically, corresponding to choices (I) and (II) in Section \ref{sec:QeR-proc} and defining
$$\mathcal{C}_{i,1}=\sum_{j=1}^M (m_{x}^{[j]}A_{ji}^T(Q^{[j]})^{-1}A_{ji}
    +m_{y}^{[j]}C_{ji}^T(R^{[j]})^{-1}C_{ji})$$
we have, respectively,
\begin{enumerate}[I)]
\item $\mathcal{P}_{1,t-1}^{[i]}=($diag$(\frac{1}{3}\Pi^{[i]}_{t-N/t-1},\frac{1}{2}\Pi^{[i]}_{t-N+1/t-1},\dots,\frac{1}{2}\Pi^{[i]}_{t-1/t-1}))^{-1}$.
    \item
    ${\mathcal{P}}^{[i]}_{1,t-1}=$diag$((\Pi^{[i]}_{t-N/t-1})^{-1}+
    \mathcal{C}_{i,1},\mathcal{C}_{i,1},\dots,\mathcal{C}_{i,1})$.
\end{enumerate}
where $m^{[i]}_x$ and $m^{[i]}_y$, $i=1,\dots,M$, are defined as the cardinalities of $\mathcal{V}^{[i]}_{x}=\{j\neq i\in\mathcal{V}:A_{ij}\neq 0\}$ and $\mathcal{V}^{[i]}_{y}=\{j\neq i\in\mathcal{V}:C_{ij}\neq 0\}$, respectively.
It is easy to show that that a solution to~\eqref{lmi:PMHE1} in case (I) always exists (recall that, at instant $t$, $\{\Pi^{[i]}_{k/t}\}_{k=t-N+1}^t$ are unknowns). However, boundedness of the solution as $t\rightarrow\infty$ is not guaranteed. This is critical because, as discussed later in the paper, boundedness of $\Pi_{k/t-1}$ is needed to prove convergence of the state estimator.\\
On the other hand, a solution to~\eqref{lmi:PMHE1} in case (II) is not always guaranteed to exist.
The following lemma addressed this issue.
\begin{lemma}
\label{lemma:suff_cond_GFT_MF}
Assume that the noise covariances are chosen as in point (II) of Section~\ref{sec:QeR-proc}. Then, a sufficient condition for~\eqref{lmi:PMHE1} to hold is that the matrix
$$\mathcal{M}_{i,1}=C_{2,i}^T(\mathbf{Q}_{N-1}^{[i]})^{-1}C_{2,i}+C_{3,i}^T(\mathbf{R}_{N-1}^{[i]})^{-1}C_{3,i}-I_{N\times N}\otimes\mathcal{C}_{i,1}$$
is positive semidefinite, where $\otimes$ denotes the Kronecker product operator. In this case~\eqref{lmi:PMHE1} is fulfilled by the update~\eqref{lmi:Kalman_UP1}.
\end{lemma}
\textbf{Proof of Lemma \ref{lemma:suff_cond_GFT_MF}}
According to \eqref{eq:unc_arr_cost_ind_PMHE1_various}
$$\begin{array}{lcl}W_{1,t}^{[i]}&=&C_{4,i}^T\text{diag}(({\Pi}_{t-N/t-1}^{*[i]})^{-1},(\mathbf{Q}_{N-1}^{[i]})^{-1},(\mathbf{R}_{N-1}^{[i]})^{-1})C_{4,i}\\
&=&C_{1,i}^T({\Pi}_{t-N/t-1}^{*[i]})^{-1}C_{1,i}+C_{2,i}^T(\mathbf{Q}_{N-1}^{[i]})^{-1}C_{2,i}+C_{3,i}^T(\mathbf{R}_{N-1}^{[i]})^{-1}C_{3,i}
\end{array}$$
where
$$C_{1,i}^T({\Pi}_{t-N/t-1}^{*[i]})^{-1}C_{1,i}=
\text{diag}(\mathcal{R}^+(\mathcal{R}(\Pi^{[i]}_{t-N/t-1},C^{[i]},
R^{[i]}_{t-N/t-1}),A^{[i]},Q^{[i]}_{t-N/t-1})^{-1},0_{n\cdot(N-1)\times n\cdot(N-1)})$$
Since choice (II) in Section~\ref{sec:QeR-proc} is done, $Q^{[i]}_{t-k/t-1}=Q^{[i]}$ and $R^{[i]}_{t-k/t-1}=R^{[i]}$ for all $k=t-N,\dots,t-1$ and ${\mathcal{P}}^{[i]}_{1,t}$ is
$${\mathcal{P}}^{[i]}_{1,t}=I_{N\times N}\otimes\mathcal{C}_{i,1}+
\text{diag}((\Pi^{[i]}_{t-N+1/t})^{-1},0_{n\cdot(N-1)\times n\cdot(N-1)})$$
Therefore, \eqref{lmi:PMHE1} is equivalent to
\begin{equation}\begin{array}{c}\text{diag}(\mathcal{R}^+(\mathcal{R}(\Pi^{[i]}_{t-N/t-1},C^{[i]},R^{[i]}_{t-N/t-1}),A^{[i]},Q^{[i]}_{t-N/t-1})^{-1},0_{n\cdot(N-1)\times n\cdot(N-1)})+C_{2,i}^T(\mathbf{Q}_{N-1}^{[i]})^{-1}C_{2,i}+C_{3,i}^T(\mathbf{R}_{N-1}^{[i]})^{-1}C_{3,i}\geq\\ I_{N\times N}\otimes\mathcal{C}_{i,1}+
\text{diag}((\Pi^{[i]}_{t-N+1/t})^{-1},0_{n\cdot(N-1)\times n\cdot(N-1)})\end{array}\label{eq:PMHE1lemmaproof}\end{equation}
If $\mathcal{M}_{i,1}>0$, then \eqref{eq:PMHE1lemmaproof} is verified if also
$$\text{diag}(\mathcal{R}^+(\mathcal{R}(\Pi^{[i]}_{t-N/t-1},C^{[i]},R^{[i]}_{t-N/t-1}),A^{[i]},Q^{[i]}_{t-N/t-1})^{-1},0_{n\cdot(N-1)\times n\cdot(N-1)})\geq
\text{diag}((\Pi^{[i]}_{t-N+1/t})^{-1},0_{n\cdot(N-1)\times n\cdot(N-1)})$$
where the equality is verified under the update~\eqref{lmi:Kalman_UP1}.\hfill$\blacksquare$\\\\
A few comments to Lemma~\ref{lemma:suff_cond_GFT_MF} are in order. First, the matrix $\mathcal{M}_{i,1}$ depends on quantities related to agents $i$ and the subsystems $j$ such that $i\in\mathcal{N}_j$. Therefore the tests $\mathcal{M}_{i,1}\geq 0$, $i=1,\dots,M$, can be done in parallel. Furthermore, diag$(\mathcal{M}_{1,1},\dots,\mathcal{M}_{M,1})\geq 0$ can be viewed as an LMI in the unknowns $(Q^{[i]})^{-1}$, $(R^{[i]})^{-1}$, $i=1,\dots,M$, and, provided that a solution exists, it defines a centralized method for the off-line computation of $Q^{[i]}$, $R^{[i]}$, $i=1,\dots,M$, guaranteeing boundedness of sequences $\{\Pi^{[i]}_{t-N/t-1}\}_{t=N}^{+\infty}$, $i=1,\dots,M$. Note also that, for fixed $Q^{[i]}$ and $R^{[i]}$, the condition $\mathcal{M}_{i,1}\geq 0$ is always fulfilled in absence of coupling, i.e., $\tilde{\mathbf{A}}=0$ and $\tilde{\mathbf{C}}=0$. Indeed, $\mathcal{M}_{i,1}\geq 0$ is equivalent to state sufficiently small couplings among neighboring subsystems.
\subsubsection{Update of $\Pi_{t-N/t-1}^{[i]}$ for PMHE2}
\label{sec:3.4.2}
The following matrix inequality must be satisfied, for each subsystem:
\begin{align}
\mathcal{P}_{2,t-1}^{[i]}\leq W_{2,t-1}^{[i]}
\label{lmi:PMHE2}
\end{align}
The matrix $W_{2,t-1}^{[i]}$ is computed, based on locally available information, as follows:
\begin{subequations}
\begin{align}\label{eq:W_PMHE2}W^{[i]}_{2,t}=(\mathcal{R}(\Pi^{*[i]}_{t-N/t-1},\mathcal{O}^{[i]}_{N-1},\Pi^{**[i]}_{t-N/t-1}))^{-1}\end{align}
where $\mathcal{O}^{[i]}_{N-1}$ is the $i$-th subsystem extended observability matrix $\mathcal{O}^{[i]}_{k}=[({C}^{[i]})^T\,\dots\,({C}^{[i]}{A}^{[i]\,\,(k-1)})^{T}]^T$, $\Pi^{*[i]}_{t-N/t-1}$ is given in \eqref{eq:unc_arr_cost_ind_PMHE1_4}, and
\begin{align}
{\Pi}_{t-N/t-1}^{**[i]}=\mathcal{R}^+(\mathbf{Q}^{[i]}_{N-1},\mathcal{C}^{[i]}_{w,N-1},\mathbf{R}^{[i]}_{N-1})
\label{eq:unc_arr_cost_ind_PMHE2_4}
\end{align}
where
\begin{align}
            \mathcal{C}^{[i]}_{w,N-1}=&\begin{bmatrix}0&0&\dots&0&0\\
            C^{[i]}& 0&\dots&0&0\\
            \vdots&\vdots&\ddots&\vdots&\vdots\\
            C^{[i]}(A^{[i]})^{N-3}& C^{[i]}(A^{[i]})^{N-4}&\dots&C^{[i]}&0
            \end{bmatrix}\label{eq:matr-conv_CwN_i}\end{align}
\end{subequations}
The matrix $\mathcal{P}_{2,t-1}^{[i]}$ has different formulations depending on which choice (i.e., whether (I) or (II) in Section~\ref{sec:QeR-proc}) has been done. We have, respectively,
\begin{enumerate}[I)]
\item $\mathcal{P}_{2,t-1}^{[i]}=(\frac{1}{2N+1}\Pi^{[i]}_{t-N/t-1})^{-1}$.
    \item
    ${\mathcal{P}}^{[i]}_{2,t-1}=\Omega_i+(\Pi^{[i]}_{t-N/t-1})^{-1}$,
    where $\Omega_i$ can be computed off-line as
    \begin{align}\Omega_i=\sum_{k=t-N}^{t-1}\|\mathbf{A}^{k-(t-N)}\|^2_2\|\tilde{\mathbf{C}}^T(\mathbf{R}^{\sharp})^{-1}\tilde{\mathbf{C}}+\tilde{\mathbf{A}}^T
     \mathbf{Q}^{-1}\tilde{\mathbf{A}}\|_2 I_{n_i}\label{eq:Omega_i_conservative}\end{align}
\end{enumerate}

Similarly to the results in Section~\ref{sec:3.4.1}, a solution to~\eqref{lmi:PMHE2} in case (I) always exists, but boundedness of the solution as $t\rightarrow\infty$ is not guaranteed. Note that boundedness of $\Pi_{k/t-1}$ will be needed to prove convergence of the state estimator.
On the other hand, a solution to~\eqref{lmi:PMHE2} in case (II) is not always guaranteed to exist. A sufficient condition providing a simple update of the initial penalty guaranteeing boundedness of the covariance matrices $\Pi_{t-N/t-1}$ can be devised, analogously to Lemma \ref{lemma:suff_cond_GFT_MF} in the case of PMHE1. In this report, however, the focus has been given to PMHE1, and future work will be devoted to a more in-depth study of PMHE2 and its properties.

\section{Convergence properties of the proposed estimators}
\label{sec:properties}
In this section
the convergence results reported in~\cite{Rao01} and~\cite{ABB03} for centralized estimators (corresponding to the trivial partition) are
extended to the proposed PMHE methods in presence of constraints. Similarly to~\cite{Rao01}, these properties are analyzed in a deterministic setting.
\begin{definition}
\label{def:convergence} Let $\Sigma$ be system
\eqref{eq:state_eq_0} with $\mathbf{w}_t=0$ and denote by $\mathbf{x}_{\Sigma}(t,\mathbf{x}_0)$
the state reached by $\Sigma$ at time $t$ starting from initial
condition $\mathbf{x}_0$. Assume that the trajectory $\mathbf{x}_{\Sigma}(t,\mathbf{x}_0)$ is
feasible, \emph{i.e.,} $\mathbf{x}_{\Sigma}(t,\mathbf{x}_0)\in \mathbb{X}$ for all
$t$. PMHE is \emph{convergent} if
$\|\hat{\mathbf{x}}_{t/t}-\mathbf{x}_{\Sigma}(t,\mathbf{x}_0)\|\stackrel{t\rightarrow \infty}{\longrightarrow} 0$.
\end{definition}
Note that, as in~\cite{Rao01}, convergence is defined assuming
that the model generating the data is noiseless, but the possible
presence of noise is taken into account in the state estimation
algorithm.
The estimation error is defined as $\mathbf{\varepsilon}_{k_1/k_2}=\mathbf{x}_{\Sigma}(k_1,\mathbf{x}_0)-\hat{\mathbf{x}}_{k_1/k_2}$.

Let $\mathcal{O}^*_{N}=$diag$\left(\mathcal{O}^{[1]}_{N},\dots,\mathcal{O}^{[M]}_{N}\right)$ be the extended observability matrix of the pair $(\mathbf{A}^*,\mathbf{C}^*)$. Furthermore, let $\mathcal{O}_{N}=[(\mathbf{C}^{\sharp})^T\,\dots\,(\mathbf{C}^{\sharp}\mathbf{A}^{N-1})^T]^T$ be the extended observability matrix of the pair $(\mathbf{A},\mathbf{C}^{\sharp})$. Denote by $f_{min}=\sigma_{min}(\mathcal{O}^*_{N})$ and $f_{max}=\sigma_{max}(\mathcal{O}^*_{N})$, the minimum and the maximum singular value of $\mathcal{O}^*_{N}$, respectively.
By Assumption~\ref{ass:local-observability}, if $N\geq \bar{n}^o$, then rank$(\mathcal{O}^{[i]}_{N})=n_i$ for all $i\in\mathcal{V}$. From this it follows that rank$(\mathcal{O}^{*}_{N})=n$, and therefore $f_{min}>0$.
Furthermore, define $\Delta_f=\|\mathcal{O}^*_{N}-\mathcal{O}_{N}\|_2$, $\kappa=\|\mathbf{A}\|_2$, and $\kappa^*=\|\mathbf{A}^*\|_2$.
\begin{lemma}
\label{def:lemma} For $N\geq\bar{n}^o$, if matrices $\Pi^{[i]}_{k/t-1}$, for $k=t-N,\dots,t-1$, satisfy \eqref{lmi:PMHE1} and are bounded, then there exist asymptotically vanishing sequences $\alpha^j_t$ (i.e.,
$\|\alpha_{t}^j\|\stackrel{t\rightarrow \infty}{\longrightarrow} 0$, $j=1,2$) such that the dynamics of the state estimation error generated by PMHE1 is given by
\begin{subequations}
\begin{align}\mathcal{O}^*_{N}{\mathbf{\varepsilon}}_{t-N/t}&=-\mathcal{C}_{N}E_{t-1}+\alpha^1_t\label{eq:step05lemma1}\\
E_{t}-M_1{\mathbf{\varepsilon}}_{t-N/t}&=M_2 E_{t-1}+\alpha^2_t\label{eq:step05lemma2}
\end{align}\label{eq:step05lemma}\end{subequations}
where $E_{t}=(\mathbf{\varepsilon}_{t-N+1/t},\dots,\mathbf{\varepsilon}_{t/t})$ and
\begin{subequations}
\label{eq:matr-conv}
\begin{align}\mathcal{C}_{N}=&\begin{bmatrix}\tilde{\mathbf{C}}&0&\dots&0\\
            \mathbf{C}^*\tilde{\mathbf{A}}& \tilde{\mathbf{C}}&\dots&0\\
            \vdots&\vdots&\ddots&\vdots\\
            \mathbf{C}^*(\mathbf{A}^*)^{N-2}\tilde{\mathbf{A}}& \mathbf{C}^*(\mathbf{A}^*)^{N-3}\tilde{\mathbf{A}}&\dots&\tilde{\mathbf{C}}
            \end{bmatrix}\label{eq:matr-conv_CN}\end{align}
\begin{align}M_1=\begin{bmatrix}
            \mathbf{A}^*\\
            \vdots\\(\mathbf{A}^*)^{N-1}\\(\mathbf{A}^*)^{N}
            \end{bmatrix}\text{, } M_2=\begin{bmatrix}
            \tilde{\mathbf{A}}& 0&\dots&0\\
            \vdots&\vdots&\ddots&\vdots\\
            (\mathbf{A}^*)^{N-2}\tilde{\mathbf{A}}& (\mathbf{A}^*)^{N-1}\tilde{\mathbf{A}}&\dots&0\\
            (\mathbf{A}^*)^{N-1}\tilde{\mathbf{A}}& (\mathbf{A}^*)^{N-2}\tilde{\mathbf{A}}&\dots&\tilde{\mathbf{A}}
            \end{bmatrix}\end{align}
\end{subequations}
\end{lemma}
\begin{lemma}
\label{def:lemmaS2} For $N\geq\bar{n}^o$, if matrices $\Pi^{[i]}_{t-N/t-1}$ satisfy \eqref{lmi:PMHE2}, are bounded, then
there exists an asymptotically vanishing sequence $\alpha_t$ (i.e.,
$\|\alpha_{t}\|\stackrel{t\rightarrow \infty}{\longrightarrow} 0$) such that the dynamics of the state estimation error generated by PMHE2 is given by
\begin{align}\mathcal{O}^*_{N}{\mathbf{\varepsilon}}_{t-N/t}
&=\left(\mathcal{O}^*_{N}-\mathcal{O}_{N}\right)\mathbf{A}{\mathbf{\varepsilon}}_{t-N-1/t-1}+\alpha_t\label{eq:lemmaS2}
\end{align}
\end{lemma}
\begin{lemma}
\label{def:lemmaS3} Assume that $\mathbf{A}^*$ is non singular, and that one of the following conditions holds:
(a) $\kappa^*< 1$ and $\mu\geq\max(0,\mu_{min})$, where $\mu_{min}=\frac{\Delta_f^2-f_{min}^2}{1-\kappa^*}$, (b) ${\kappa}^*\geq 1$ and $0\leq \mu\leq \mu_{max}=\frac{f_{min}^2-\Delta_f^2}{\kappa^{*\,2}-1}$. If $N\geq\bar{n}^o$, then
there exists an asymptotically vanishing sequence $\alpha_t$ (i.e.,
$\|\alpha_{t}\|\stackrel{t\rightarrow \infty}{\longrightarrow} 0$) such that the dynamics of the state estimation error generated by PMHE3 obeys to~\eqref{eq:lemmaS2}.
\end{lemma}
We are now in the position to state the main convergence results.
\begin{theorem}
\label{th:main}
$\,$
\begin{itemize}
\item[I)] Under the assumptions of Lemma~\ref{def:lemma}, if the matrix
$$\Phi_1=M_2-M_1\left((\mathcal{O}^*_{N})^T\mathcal{O}^*_{N}\right)^{-1}(\mathcal{O}^*_{N})^T\mathcal{C}_{N}$$
is Schur, then PMHE1 is convergent.
\item[II)] Under the assumptions of Lemma~\ref{def:lemmaS2}, if the matrix
\begin{align}
\Phi_2=\left((\mathcal{O}^*_{N})^T\mathcal{O}^*_{N}\right)^{-1}(\mathcal{O}^*_{N})^T\left(\mathcal{O}^*_{N}-\mathcal{O}_{N}\right)\mathbf{A}
\label{eq:Phi2}
\end{align}
is Schur, then PMHE2 is convergent.
\item[III)] Under the assumptions of Lemma~\ref{def:lemmaS3}, if the matrix $\Phi_2$ is
Schur then
PMHE3 is convergent.
\end{itemize}
\end{theorem}
\subsection{Simpler conditions for PMHE2 and PMHE3 and a measure \GFT{of the quality of the partition}}
We start providing sufficient conditions for the convergence to zero of the error dynamics~\eqref{eq:lemmaS2}.
To this aim, we define the scalars $\nu$ and $a_0$ as
\begin{align}
\nu=\frac{f_{max}\Delta_f}{f_{min}^2},\quad a_0=\nu\,\kappa
\label{eq:a_0}
\end{align}
\begin{theorem}
Under the assumptions of Lemma~\ref{def:lemmaS2} [resp. Lemma~\ref{def:lemmaS3}], if $a_0<1$, then PMHE2 [resp. PMHE3] is convergent.
\label{th:suff_conv_2-3}
\end{theorem}
In view of this result, convergence of PMHE2 and PMHE3 depends on the norm $\kappa$ of the transition matrix $\mathbf{A}$ (which is a function of the problem data), and the constant $\nu$, given in~\eqref{eq:a_0}, which depends on the adopted partition. Therefore, $\nu$ can be considered as a measure of the quality of the partition, for state estimation purposes. Specifically, the smaller $\nu$, the better the partition. On the other hand, $\Delta_f$ is expected to increase with the number of arcs in the graph and this results in an increase of the index $\nu$.\\\\
Concerning only PMHE3, Lemma~\ref{def:lemmaS3}, part~III of Theorem~\ref{th:main} and Theorem~\ref{th:suff_conv_2-3} generalize the results of~\cite{ABB03} to distributed estimation. Using similar arguments to~\cite{ABB03} we also obtain the following results, that provide an alternative convergence condition.
\begin{lemma}
\label{def:lemmaS3_unc} For all $i\in\mathcal{V}$, if $N\geq\bar{n}^o$,
the dynamics of the state estimation error provided by PMHE3 is given by

\begin{align}&{\mathbf{\varepsilon}}_{t-N/t}
=\left(\mu I_n+(\mathcal{O}^*_{N})^T\mathcal{O}^*_{N}\right)^{-1}\left[\mu I_n+(\mathcal{O}^*_{N})^T\left(\mathcal{O}^*_{N}-\mathcal{O}_{N}\right)\right]\mathbf{A}{\mathbf{\varepsilon}}_{t-N-1/t-1}\label{eq:lemmaS3_unc}
\end{align}
\end{lemma}
Note that the error dynamics~\eqref{eq:lemmaS3_unc} coincide with the error dynamics ${\mathbf{\varepsilon}}_{t-N/t}=\bm{\Phi}_2{\mathbf{\varepsilon}}_{t-N-1/t-1}$ when $\mu=0$. Hence Lemma~\ref{def:lemmaS3_unc} generalizes the stability condition of point (II) of Theorem~\ref{th:main}. Similarly, the stability condition of Theorem~\ref{th:suff_conv_2-3} is generalized in the following corollary.
\begin{corollary}
If $N\geq\bar{n}^o$, then PMHE3 is convergent if
\begin{align}
a(\mu)=\frac{\mu+f_{max}\Delta_f}{\mu+f_{min}^2}\kappa=\frac{\frac{\mu}{f_{min}^2}+\nu}{\frac{\mu}{f_{min}^2}+1}\kappa<1
\label{eq:amu}
\end{align}
\label{cor:suff_cond_3_unc}
\end{corollary}

\section{Algorithmic summary for PMHE1}
In this section we summarize the steps one has to take in order to apply the PMHE1 scheme. Similar algorithms can be derived for PMHE2 and PMHE3. Recalling the options discussed in Section~\ref{sec:QeR-proc}, we consider the two following cases.
\begin{enumerate}[I)]
\item \textbf{$Q^{[i]}_{k/t-1}$ and $R^{[i]}_{k/t-1}$ are chosen as in point (I) in Section~\ref{sec:QeR-proc}}\\\\
\textbf{Offline design}\\
Compute the matrices $\{\Pi^{[i]}_{k/t-1}\}_{k=t-N}^{t-1}$ from the LMI~\eqref{lmi:PMHE1} and verify that the sequence $\{\Pi^{[i]}_{k/t-1}\}_{k=t-N}^{t-1}$ is bounded for $t\rightarrow+\infty$.\\\\
\textbf{Online state estimation\\}
\begin{itemize}
\item Initialization: each subsystem $i\in\mathcal{V}$ stores the initialization matrices $\{\Pi^{[i]}_{k/N-1}\}_{k=0}^{N-1}$ and the estimates $\hat{x}^{[i]}_{k/N-1}$, $k=0,\dots,N-1$, and transmits them to all the subsystems $j$ such that $i\in\mathcal{V}^{[j]}$;\\
\item if $t\geq N$, at each time instant, each subsystem $i\in\mathcal{V}$
\begin{itemize}
\item[-] receives, from all the neighboring subsystems $j\in\mathcal{V}^{[i]}$, the estimates $\hat{x}^{[j]}_{k/t-1}$, $k=t-N,\dots,t-1$, and the matrices $\Pi^{[j]}_{k/t-1}$, $k=t-N,\dots,t-1$;
\item[-] solves the problem MHE\emph{-i}, with stage cost
    $$L_1^{[i]}=\frac{1}{2}\|\hat{w}^{[i]}_k\|_{(Q^{[i]}_{k/t-1})^{-1}}^2+
    \frac{1}{2}\|\hat{v}^{[i]}_k\|_{({R}^{[i]}_{k/t-1})^{-1}}^2 $$
    where ${Q}^{[i]}_{k/t-1}$ and ${R}^{[i]}_{k/t-1}$ are computed as in~\eqref{eq:QandR_PMHE1_Q} and~\eqref{eq:QandR_PMHE1_R}, respectively, and with initial penalty
    $$\Gamma_{1,t-N}^{[i]}=\frac{1}{2}\|\hat{x}_{t-N}^{[i]}
    -\hat{x}^{[i]}_{t-N/t-1}\|^2_{({\Pi_{t-N/t-1}^{[i]}})^{-1}}$$
\item[-] updates the matrices $\{\Pi^{[i]}_{k/t}\}_{k=t-N+1}^{t}$ according to the LMI~\eqref{lmi:PMHE1};
\item[-] broadcasts the matrices $\{\Pi^{[i]}_{k/t}\}_{k=t-N+1}^{t}$ and the estimates $\hat{x}^{[i]}_{k/t}$, $k=t-N+1,\dots,t$ to all the subsystems $j$ such that $i\in\mathcal{V}^{[j]}$.\\
\end{itemize}
\end{itemize}
\item \textbf{$Q^{[i]}_{k/t-1}$ and $R^{[i]}_{k/t-1}$ are chosen as in point (II) in Section~\ref{sec:QeR-proc}}\\\\
\textbf{Offline design}\\
Choose $Q^{[i]}$ and $R^{[i]}$ such that the LMIs $\mathcal{M}_{i,1}\geq 0$ are fulfilled for all $i=1,\dots,M$.\\\\
\textbf{Online state estimation}\\
\begin{itemize}
\item Initialization: each subsystem $i\in\mathcal{V}$ stores the initialization matrix $\Pi^{[i]}_{0/N-1}$ and the estimates $\hat{x}^{[i]}_{k/N-1}$, $k=0,\dots,N-1$, and transmits $\hat{x}^{[i]}_{k/N-1}$, $k=0,\dots,N-1$ to all the subsystems $j$ such that $i\in\mathcal{V}^{[j]}$;\\
\item if $t\geq N$, at each time instant, each subsystem $i\in\mathcal{V}$
\begin{itemize}

\item[-] receives, from all the neighboring subsystems $j\in\mathcal{V}^{[i]}$, the estimates $\hat{x}^{[j]}_{k/t-1}$, $k=t-N,\dots,t-1$;
\item[-] solves the problem MHE\emph{-i}, with stage cost
    $$L_1^{[i]}=\frac{1}{2}\|\hat{w}^{[i]}_k\|_{(Q^{[i]})^{-1}}^2+
    \frac{1}{2}\|\hat{v}^{[i]}_k\|_{({R}^{[i]})^{-1}}^2 $$
    and with initial penalty
    $$\Gamma_{1,t-N}^{[i]}=\frac{1}{2}\|\hat{x}_{t-N}^{[i]}
    -\hat{x}^{[i]}_{t-N/t-1}\|^2_{({\Pi_{t-N/t-1}^{[i]}})^{-1}}$$
\item[-] updates the matrix $\Pi^{[i]}_{t-N+1/t}$ according to~\eqref{lmi:Kalman_UP1};
\item[-] broadcasts the estimates $\hat{x}^{[i]}_{k/t}$, $k=t-N+1,\dots,t$ to all the subsystems $j$ such that $i\in\mathcal{V}^{[j]}$.
\end{itemize}
\end{itemize}
\end{enumerate}
\section{Discussion and conclusions}
\label{sec:discussion}
In this report three distributed state-estimation algorithms, namely PMHE1, PMHE2 and PMHE3, for partitioned large-scale systems have been proposed and sufficient conditions for convergence have been established. The three solutions have different features in terms of communication requirements among subsystems, accuracy and computational complexity. More specifically, PMHE1 \MF{relies on a partially connected communication graph} in the sense that subsystems exploit a communication network where links are present only if subsystem dynamics are coupled. Algorithms PMHE2 and PMHE3 assume an all-to-all communication but a reduced amount of information is transmitted over each communication channel.
The main difference between PMHE1 and PMHE2 consists in the type of communication required among subsystems, and on how the estimates of $u^{[i],x}_t$ and $u^{[i],y}_t$ are used. While in PMHE1 and PMHE2 the transmitted information amounts to state estimates and estimation error covariances, in PMHE3 no information on the noise variances is required and the weights on the different components of the cost functions are constant. This allows for a significant reduction in terms of transmission and computational load, at the price of a loss in noise filtering performance.
\section*{Acknowledgements}
This research has been supported by
the European 7th framework STREP projects ``Hierarchical
and distributed model predictive control (HD-MPC)", contract
number INFSO-ICT-223854, and ``HYCON2 Network of excellence" [FP7/2007-2013], under grant agreement n. 257462 .

\bibliographystyle{plain}
\bibliography{DMHE_bib_GFT2}

\appendix
\section{Optimization}
\label{app:optimization}
In this appendix we recall some known results which are used throughout the paper.
In Appendix~\ref{sec:PMHE1-2} that follow, the following results are used (see \cite{RawMayneBook} for a proof).
\begin{lemma}
\label{lemma:opt - smoothing}
Let
$$V(x)=\frac{1}{2}\|x-\hat{x}\|^2_{P^{-1}}+\frac{1}{2}\|y-Cx\|^2_{R^{-1}}$$
Then
$$V(x)=\frac{1}{2}\|x-x^*\|^2_{H^{-1}}+V(x^*)$$
where
$$\begin{array}{lcl}H&=&\mathcal{R}(P,C,R)\\
x^*&=&\hat{x}+\mathcal{K}(P,C,R)(y-C\hat{x})
\end{array}$$
\end{lemma}
\begin{lemma}
\label{lemma:opt - partial update}
Let
$$\begin{array}{rcl}V^-(x(0),x(1))&=&\frac{1}{2}\|x(0)-\hat{x}(0)\|^2_{P(0)^{-1}}+\frac{1}{2}\|x(1)-Ax(0)\|^2_{Q^{-1}}\\
V^{-*}(\hat{x}(1))&=&\min_{x(0)}\{V^-(x(0),x(1))\text{\emph{ subject to }} x(1)=\hat{x}(1)\}
\end{array}$$
Then
$$\begin{array}{rcl}V^{-*}(\hat{x}(1))&=&\frac{1}{2}\|\hat{x}(1)-A\hat{x}(0)\|^2_{P(1)^{-1}}\\
P(1)&=&\mathcal{R}^+(P(0),A,Q)\\
x(0)^*&=&\text{\emph{arg}}\min_{x(0)}\{V^-(x(0),x(1))\text{\emph{ subject to }} x(1)=\hat{x}(1)\}\\
&=&\hat{x}(0)+\mathcal{K}(P(0),A,Q)(\hat{x}(1)-A\hat{x}(0))
\end{array}$$
\end{lemma}
\begin{lemma}
\label{lemma:opt - centering}
Let \emph{rank}$(C)=$\emph{dim}$(\xi)$, then
$$\frac{1}{2}\|C x - \xi\|^2_{W^{-1}} = \frac{1}{2}\|x-\hat{x}\|^2_{C^T W^{-1}C}$$
where
$$\begin{array}{lcl}
\hat{x}&=&(C^TW^{-1}C)^{-1}C^TW^{-1}\xi
\end{array}$$
\end{lemma}
The next Lemma is a generalization of Lemma~11 in \cite{Rao01}
\begin{lemma}
\label{lemma:Bound}
Let $\vartheta(z)=\|z-\xi\|_Q^2+d$ where $z$, $\xi\in\mathbb{R}^n$, $d\in\mathbb{R}$ and the matrix $Q$ is symmetric and positive definite. Let $\Gamma$ and $\Omega$ be nonempty closed and convex sets verifying $\Gamma\subseteq\Omega$. Let $\hat{z}$ and $\bar{z}$ be solutions to the following quadratic programs
$$\vartheta(\hat{z})=\min_{z\in\Omega}\vartheta(z),\,\,\vartheta(\bar{z})=\min_{z\in\Gamma}\vartheta(z)$$
and define $\Delta z=\bar{z}-\hat{z}$. Then, one has
\begin{equation}
\vartheta(\bar{z})\geq \vartheta(\hat{z})+\vartheta(\Delta z+\xi)-d
\label{eq:Bound_Rawlings_Rao}
\end{equation}
\end{lemma}
\textbf{Proof of Lemma \ref{lemma:Bound}}\\
One has
\begin{align}\vartheta(\bar{z})&=\vartheta(\hat{z}+\Delta z)=\|\hat{z}+\Delta z-\xi\|_Q^2+d\nonumber\\
&=\|\hat{z}-\xi\|_Q^2+\|\Delta z\|_Q^2+2(\hat{z}-\xi)^TQ\Delta z+d\label{eq:Bound_4}
\end{align}
where $\|\hat{z}-\xi\|_Q^2=\vartheta(\hat{z})-d$, $\|\Delta z\|_Q^2=\vartheta(\Delta z+\xi)-d$ and, since $\nabla_z\vartheta(\hat{z})=2 Q (\hat{z}-\xi)$, then
$$2(\hat{z}-\xi)^TQ\Delta z = \nabla_z\vartheta(\hat{z})^T\Delta z$$
By optimality of $\hat{z}$ it holds that
$$\nabla_z\vartheta(\hat{z})^T(z-\hat{z})\geq 0\,\forall z\in\Omega$$
and then
$$\nabla_z\vartheta(\hat{z})^T\Delta z \geq 0\,\text{ with }\Delta z=z-\hat{z},\forall z\in\Gamma$$
Therefore one has that, from \eqref{eq:Bound_4}
$$\vartheta (\bar{z})\geq \vartheta(\hat{z})-d+\vartheta(\Delta z+\xi)-d+d$$
which corresponds with \eqref{eq:Bound_Rawlings_Rao}.
\section{Proofs}
\label{sec:PMHE1-2}
Denote by $\mathbf{J}_r$ the sum (for all the $M$ subsystems) of the local cost functions $J_r^{[i]}$, given by
(\ref{eq:i-th cost fcn})
\begin{equation}\mathbf{J}_r=\sum_{i=1}^M J_r^{[i]}(t-N,t,\hat{x}^{[i]}_{t-N},\hat{w}^{[i]},\hat{v}^{[i]},\Gamma^{[i]}_{r,t-N})
\label{eq:tot cost fcn}\end{equation}
Define the collective vectors
${\hat{\mathbf{x}}}_t=(\hat{x}^{[1]}_t,\dots,\hat{x}^{[M]}_t)\in\mathbb{R}^{n}$,
$\hat{\mathbf{v}}_t=(\hat{v}^{[1]}_t,\dots,\hat{v}^{[M]}_t)\in\mathbb{R}^{\bar{p}}$, ${\hat{\mathbf{w}}}_t=(\hat{w}^{[1]}_t,\dots,\hat{w}^{[M]}_t)\in\mathbb{R}^{n}$
and rewrite $\mathbf{J}_r$ as
\begin{equation}\begin{array}{l}\mathbf{J}_r
=\sum_{k=t-N}^{t-1}\mathbf{L}_r({\hat{\mathbf{w}}}_k, \hat{\mathbf{v}}_k)+\bm{\Gamma}_{r,t-N}({\hat{\mathbf{x}}}_{t-N};{\hat{\mathbf{x}}}_{t-N/t-1})
\end{array}\label{eq:tot_cost_fcn_2}\end{equation}
For $r=1,2$:
\small\begin{align}
\mathbf{L}_r
=&\frac{1}{2}\|\hat{\mathbf{v}}_k\|_{(\mathbf{R}_{k/t-1})^{-1}}^2 + \frac{1}{2}\|{\hat{\mathbf{w}}}_k\|_{(\mathbf{Q}_{k/t-1})^{-1}}^2\label{eq:tot_step_term}\\
\bm{\Gamma}_{r,t-N}
=&\frac{1}{2}\|{\hat{\mathbf{x}}}_{t-N}-{\hat{\mathbf{x}}}_{t-N/t-1}\|^2_{\bm{\Pi}_{t-N/t-1}^{-1}}+\bm{\Theta}^*_{t-1}
\label{eq:tot_pen_term}\end{align}\normalsize
where $\mathbf{R}_{k/t-1}=$diag$(R^{[1]}_{k/t-1},\dots,R^{[M]}_{k/t-1})$, $\mathbf{Q}_{k/t-1}=$
diag$(Q^{[1]}_{k/t-1},\dots,Q^{[M]}_{k/t-1})$ and $\bm{\Theta}^*_{t-1}=\sum_{i=1}^M{\Theta}^{*[i]}_{r,t-1}$.
On the other hand, for PMHE3:
\begin{align}
\mathbf{L}_3
=&\frac{1}{2}\|\hat{\mathbf{v}}_k\|^2\label{eq:tot_step_term3}\\
\bm{\Gamma}_{3,t-N}
=&\frac{\mu}{2}\|{\hat{\mathbf{x}}}_{t-N}-{\hat{\mathbf{x}}}_{t-N/t-1}\|^2+\bm{\Theta}^*_{3,t-1}
\label{eq:tot_pen_term3}\end{align}
Constraints~\eqref{eq:update_x&v} can be written in the following collective form
\begin{subequations}\begin{align}
&\hat{\mathbf{x}}_{k+1}=\mathbf{A}^*\,\hat{\mathbf{x}}_{k}+\tilde{\mathbf{A}}\,\tilde{\mathbf{x}}_{k/t-1}+\hat{\mathbf{w}}_k\label{eq:collective-update}\\
&\mathbf{y}^{\sharp}_k=\mathbf{C}^*\,\hat{\mathbf{x}}_{k}+\tilde{\mathbf{C}}\,\tilde{\mathbf{x}}_{k/t-1}+\hat{\mathbf{v}}_k\label{eq:collective-out}\\
&\hat{\mathbf{x}}_k\in \mathbb{X}\text{ if } r=1,2\\
&\hat{\mathbf{w}}_k=0\text{ if }r=3\label{eq:collective constraints_w_for_PMHE3}
\end{align}\label{eq:collective constraints}\end{subequations}
with $k=t-N,\dots,t-1$. Notice that $\tilde{\mathbf{x}}_{k/t-1}$, $k=t-N,\dots,t-1$, depend on the adopted transmission model. Namely, in case of neighbor-to-neighbor transmission (Model 1, and therefore in case the PMHE1 scheme is used), $\tilde{\mathbf{x}}_{k/t-1}=\hat{\mathbf{x}}_{k/t-1}$ for all $k=t-N,\dots,t-1$, while in case of all-to-all transmission (Model 2, and therefore in case the PMHE2 scheme is used), $\tilde{\mathbf{x}}_{k/t-1}=\mathbf{A}^{k-(t-N)}\hat{\mathbf{x}}_{t-N/t-1}$ for all $k=t-N,\dots,t-1$.
The solution to
\begin{align}
\min_{\hat{\mathbf{x}}_{t-N},\{\hat{\mathbf{w}}_k\}_{k=t-N}^{t-1}}\mathbf{J}_r(t-N,t,\hat{\mathbf{x}}_{t-N},\hat{\mathbf{w}},\hat{\mathbf{v}},\bm{\Gamma}_{r,t-N})
\label{eq:collective_min_pb}
\end{align}
where $\hat{\mathbf{w}}$ and $\hat{\mathbf{v}}^{[i]}$ are shorthand notation for $\{\hat{\mathbf{w}}_k\}_{k=t-N}^{t-1}$ and $\{\hat{\mathbf{v}}_k\}_{k=t-N}^{t-1}$ respectively, is equivalent to the solution to the MHE-$i$ problems~\eqref{eq:min-pb}, in the sense that $\hat{x}^{[i]}_{t-N/t},\{\hat{w}^{[i]}_{k/t}\}_{k=t-N}^{t-1}$ is a solution to~\eqref{eq:min-pb} if and only if
$\hat{\mathbf{x}}_{t-N/t},\{\hat{\mathbf{w}}_{k/t}\}_{k=t-N}^{t-1}$ is a solution to~\eqref{eq:collective_min_pb},
where $\hat{\mathbf{w}}_{k/t}=(\hat{w}^{[1]}_{k/t},\dots,\hat{w}^{[M]}_{k/t})$. In fact, at time $t$, variables $\tilde{\mathbf{x}}_{k/t-1}$ are fixed inputs for the system~\eqref{eq:collective constraints}.

We define the \emph{transit cost} for the proposed PMHE schemes. For PMHEr, the local transit cost in a generic sequence $z^{[i]}_k\in \mathbb{R}^{n_i}$ $k=k_1,\dots,k_2$, for subsystem $i$ is defined as
\begin{equation}
\begin{array}{c}\Xi_{r,[k_1:k_2]/t}^{[i]}(\{z^{[i]}_k\}_{k=k_1}^{k_2})=\min_{\hat{x}_{t-N}^{[i]},\hat{w}^{[i]}}\left\{J^{[i]}_r(t-N,t,\hat{x}^{[i]}_{t-N},\hat{w}^{[i]},\hat{v}^{[i]}
,\Gamma^{[i]}_{r,t-N})\right.\\ \left.\text{ subject to~\eqref{eq:update_x&v}  and } {\hat{x}}^{[i]}_{k}
=z^{[i]}_k\text{ for }k=k_1,\dots,k_2\right\}
\end{array}
\label{eq:arrival_cost_def_local}
\end{equation}
The collective transit cost in a generic sequence $\mathbf{z}_k=(z^{[1]}_k,\dots,z^{[M]}_k)\in \mathbb{R}^{n}$ $k=k_1,\dots,k_2$, is defined as
\begin{equation}
\begin{array}{c}
\bm{\Xi}_{r,[k_1:k_2]/t}(\{\mathbf{z}_k\}_{k=k_1}^{k_2})=\min_{\hat{\mathbf{x}}_{t-N},\hat{\mathbf{w}}}
\left\{\mathbf{J}_r(t-N,t,{\hat{\mathbf{x}}}_{t-N},\hat{\mathbf{w}},\hat{\mathbf{v}},\bm{\Gamma}_{r,t-N}) \right.\\\left.\text{ subject to }\eqref{eq:collective constraints}\text{ and } {\hat{\mathbf{x}}}_{k}
=\mathbf{z}_k\text{ for }k=k_1,\dots,k_2\right\}\end{array}
\label{eq:arrival_cost_def}
\end{equation}
and it holds that
\begin{align}
\bm{\Xi}_{r,[k_1:k_2]/t}(\{\mathbf{z}_k\}_{k=k_1}^{k_2})=\sum_{i=1}^M \Xi_{r,[k_1:k_2]/t}^{[i]}(\{z^{[i]}_k\}_{k=k_1}^{k_2})
\label{eq:arrival_cost_sum}
\end{align}

In case of PMHE1, the key condition to prove Lemma~~\ref{def:lemma} is that, for all sequences $\mathbf{z}_k\in\mathbb{X}$, $k=t-N,\dots,t-1$
\begin{subequations}
\label{eq:assumption-penalty}
\begin{align}
&\sum_{k=t-N}^{t-1}\frac{1}{2}\|\mathbf{z}_k-\hat{\mathbf{x}}_{k/t-1}\|_{
    \tilde{\mathbf{C}}^T(\mathbf{R}_{k/t-1})^{-1}\tilde{\mathbf{C}}+\tilde{\mathbf{A}}^T(\mathbf{Q}_{k/t-1})^{-1}\tilde{\mathbf{A}}
    }^2 +
    \frac{1}{2}\|\mathbf{z}_{t-N}-\hat{\mathbf{x}}_{t-N/t-1}\|^2_{\bm{\Pi}_{t-N/t-1}^{-1}}+\bm{\Theta}^*_{t-1}\leq
\bm{\Xi}_{1,[t-N:t-1]/t-1}(\{\mathbf{z}_k\}_{k=t-N}^{t-1}) \label{eq:assumption-penalty_PMHE1}\end{align}
For PMHE2 we will require that, for all $\mathbf{z}_{t-N}\in\mathbb{X}$
\begin{align}
&\frac{1}{2}\|\mathbf{z}_{t-N}-\hat{\mathbf{x}}_{t-N/t-1}\|_{    \sum_{k=t-N}^{t-1}(\mathbf{A}^{k-(t-N)})^T\left(\tilde{\mathbf{C}}^T(\mathbf{R}_{k/t-1})^{-1}\tilde{\mathbf{C}}+\tilde{\mathbf{A}}^T(\mathbf{Q}_{k/t-1})^{-1}\tilde{\mathbf{A}}
    \right)\mathbf{A}^{k-(t-N)}+\bm{\Pi}_{t-N/t-1}^{-1}}^2+\bm{\Theta}^*_{t-1}\leq
\bm{\Xi}_{2,t-N/t-1}(\mathbf{z}_{t-N}) \label{eq:assumption-penalty_PMHE2}\end{align}
Finally, for PMHE3 we require that, for all $\mathbf{z}_{t-N}\in\mathbb{R}^n$
\begin{align}
&\frac{1}{2}\|\mathbf{z}_{t-N}-\hat{\mathbf{x}}_{t-N/t-1}\|^2_{\left(\mathcal{O}_N-\mathcal{O}_{N}^*\right)^T
    \left(\mathcal{O}_N-\mathcal{O}_{N}^*\right)+\mu\,I_n}+\bm{\Theta}^*_{3,t-1}\leq
\bm{\Xi}_{3,t-N/t-1}(\mathbf{z}_{t-N}) \label{eq:assumption-penalty_PMHE3}\end{align}
\end{subequations}

Under~\eqref{eq:assumption-penalty} the first step towards the convergence of the PMHE estimators is the following lemma.
\begin{lemma}
\label{lemma:common_lemma}
If~\eqref{eq:assumption-penalty} holds then, for all the PMHE schemes
\begin{align}
\sum_{k=t-N}^{t-1}\mathbf{L}_r({\hat{\mathbf{w}}}_{k/t}, \hat{\mathbf{v}}_{k/t})\stackrel{t\rightarrow\infty}{\longrightarrow}0\label{eq:Lconverges}
\end{align}
\end{lemma}
The following results are specific for each estimation scheme. We first provide conditions guaranteeing~\eqref{eq:assumption-penalty} to hold, then we investigate the implications of~\eqref{eq:Lconverges}, in terms of estimation error convergence.

\begin{lemma}
\label{lemma:step_4}
For PMHE1 and PMHE2, if \eqref{lmi:PMHE1} and \eqref{lmi:PMHE2}, respectively, are satisfied then~\eqref{eq:assumption-penalty} holds.
\end{lemma}

\begin{lemma}
\label{lemma:step_4_S3}
For PMHE3, if the conditions of Lemma \ref{def:lemmaS3} are satisfied, then~\eqref{eq:assumption-penalty} holds.
\end{lemma}

\begin{lemma}
\label{lemma:proof_step_3}
Assume that $N\geq\bar{n}^o$ and equation~\eqref{eq:Lconverges} holds. Then
\begin{enumerate}[a)]
\item if $\mathbf{Q}_{k/t-1}$ and $\mathbf{R}_{k/t-1}$ are bounded for all $t$ and for all $k=t-N,\dots,t-1$, then the dynamics of the state estimation error provided by PMHE1 is given by~\eqref{eq:step05lemma};
\item if $\mathbf{Q}_{k/t-1}$ and $\mathbf{R}_{k/t-1}$ are bounded for all $t$ and for all $k=t-N,\dots,t-1$, then the dynamics of the state estimation error provided by PMHE2 is given by~\eqref{eq:lemmaS2};
    \item the dynamics of the state estimation error provided by PMHE3 is given by~\eqref{eq:lemmaS2}.
\end{enumerate}
\end{lemma}

\textbf{Proof of Lemmas~\ref{def:lemma},~\ref{def:lemmaS2} and~\ref{def:lemmaS3}}\\
For PMHE1 and PMHE2 [resp, for PMHE3], in view of Lemma~\ref{lemma:step_4} [resp. Lemma~\ref{lemma:step_4_S3}],~\eqref{eq:assumption-penalty} holds. Therefore, from Lemma~\ref{lemma:common_lemma},~\eqref{eq:Lconverges} is verified. The proofs are concluded, by resorting to Lemma~\ref{lemma:proof_step_3}. The latter can be applied under the assumption that $Q^{[i]}_{k/t-1}$ and $R^{[i]}_{k/t-1}$ are bounded, which is verified, since the matrices $\Pi^{[i]}_{k/t}$ remain bounded. \hfill$\blacksquare$\\
\textbf{Proof of Lemma~\ref{lemma:common_lemma}}\\
The proof is similar to the one in Proposition~5 in~\cite{Rao01}.
For all $t\geq 0$ and all $r=1,2,3$, one has
\begin{equation}\bm{\Theta}_{r,t}^*-\bm{\Theta}_{r,t-1}^*\geq \sum_{k=t-N}^{t-1}\mathbf{L}_r({\hat{\mathbf{w}}}_{k/t}, \hat{\mathbf{v}}_{k/t})
\label{eq:proof_step_1}\end{equation}
From now on, different derivations arise for each PMHE scheme.
\begin{itemize}
\item For PMHE1, by optimality $\bm{\Theta}_{1,t}^*\leq \bm{\Xi}_{1,[t-N+1:t]/t}(\{\mathbf{x}_{\Sigma}(k,\mathbf{x}_0)\}_{k=t-N+1}^t),\,\,\forall t$.
    Note that the trajectory for~\eqref{eq:collective-update}, stemming from $\hat{\mathbf{x}}_{t-N}=\mathbf{x}_{\Sigma}(t-N,\mathbf{x}_0)$ with $\hat{\mathbf{w}}_k=\hat{\mathbf{w}}^*_k$ and $$\hat{\mathbf{w}}^*_k=
    \tilde{\mathbf{A}}(\mathbf{x}_{\Sigma}(k,\mathbf{x}_0)-\hat{\mathbf{x}}_{k/t-1})$$
    for $k=t-N,\dots,t-1$ is $\hat{\mathbf{x}}_{k}=\mathbf{x}_{\Sigma}(k,\mathbf{x}_0)$ for $k=t-N+1,\dots,t$ and hence it is a feasible trajectory. Since $\mathbf{y}^{\sharp}_k$, $k=t-N,\dots,t-1$ corresponds to the deterministic system output from Definition~\ref{def:convergence}, it follows that $$\hat{\mathbf{v}}_k=\mathbf{y}^{\sharp}_k-(\mathbf{C}^*\mathbf{x}_{\Sigma}(k,\mathbf{x}_0)+\tilde{\mathbf{C}}\hat{\mathbf{x}}_{k/t-1})=
    \tilde{\mathbf{C}}(\mathbf{x}_{\Sigma}(k,\mathbf{x}_0)-\hat{\mathbf{x}}_{k/t-1})$$
    for all $k=t-N,\dots,t-1$.
    Moreover, by optimality
    $$\begin{array}{l}\bm{\Xi}_{1,[t-N+1:t]/t}(\{\mathbf{x}_{\Sigma}(k,\mathbf{x}_0)\}_{k=t-N+1}^t)\leq \mathbf{J}_1(t-N,t,{\mathbf{x}}_{\Sigma}(t-N,\mathbf{x}_0),\hat{\mathbf{w}},\hat{\mathbf{v}},\bm{\Gamma}_{1,t-N})\end{array}$$
    From~\eqref{eq:tot_cost_fcn_2}, one has \begin{align}&\mathbf{J}_1(t-N,t,{\mathbf{x}}_{\Sigma}(t-N,\mathbf{x}_0),\hat{\mathbf{w}},\hat{\mathbf{v}},\bm{\Gamma}_{1,t-N})=\nonumber\\
    &=\sum_{k=t-N}^{t-1}\mathbf{L}_1(\hat{\mathbf{w}}_k, \hat{\mathbf{v}}_k)+\bm{\Gamma}_{1,t-N}({\mathbf{x}}_{\Sigma}(t-N,\mathbf{x}_0);{\hat{\mathbf{x}}}_{t-N/t-1}),
            \nonumber\end{align}
    In view of~\eqref{eq:tot_step_term} and~\eqref{eq:tot_pen_term} we have that
    \begin{align}\mathbf{J}_1&(t-N,t,{\mathbf{x}}_{\Sigma}(t-N,\mathbf{x}_0),\hat{\mathbf{w}},\hat{\mathbf{v}},\bm{\Gamma}_{1,t-N})=\nonumber\\
    &=\sum_{k=t-N}^{t-1}\frac{1}{2}\|\tilde{\mathbf{C}}(\mathbf{x}_{\Sigma}(k,\mathbf{x}_0)-\hat{\mathbf{x}}_{k/t-1})\|_{(\mathbf{R}_{k/t-1})^{-1}}^2 +\frac{1}{2}\|\tilde{\mathbf{A}}(\mathbf{x}_{\Sigma}(k,\mathbf{x}_0)-\hat{\mathbf{x}}_{k/t-1})\|_{(\mathbf{Q}_{k/t-1})^{-1}}^2+
    \frac{1}{2}\|\mathbf{x}_{\Sigma}(t-N,\mathbf{x}_0)-{\hat{\mathbf{x}}}_{t-N/t-1}\|^2_{\bm{\Pi}_{t-N/t-1}^{-1}}+\bm{\Theta}^*_{1,t-1}
            \nonumber\\
    &=\sum_{k=t-N}^{t-1}\frac{1}{2}\|\mathbf{x}_{\Sigma}(k,\mathbf{x}_0)-\hat{\mathbf{x}}_{k/t-1}\|_{
    \tilde{\mathbf{C}}^T(\mathbf{R}_{k/t-1})^{-1}\tilde{\mathbf{C}}+\tilde{\mathbf{A}}^T(\mathbf{Q}_{k/t-1})^{-1}\tilde{\mathbf{A}}
    }^2 +
    \frac{1}{2}\|\mathbf{x}_{\Sigma}(t-N,\mathbf{x}_0)-{\hat{\mathbf{x}}}_{t-N/t-1}\|^2_{\bm{\Pi}_{t-N/t-1}^{-1}}+\bm{\Theta}^*_{1,t-1}
            \nonumber\end{align}
    and in view of~\eqref{eq:assumption-penalty_PMHE1},
   $\bm{\Theta}_{1,t}^*\leq \bm{\Xi}_{1,[t-N:t-1]/t-1}(\{\mathbf{x}_{\Sigma}(k,\mathbf{x}_0)\}_{k=t-N}^{t-1})$. We can iterate this procedure and prove that
   \begin{equation}\bm{\Theta}_{1,t}^*\leq {\Gamma}_{1,0}({\mathbf{x}}_{0})\label{eq:proof_step_2_1}\end{equation}
    for all $t$, for any $\mathbf{x}_{0}\in\mathbb{X}$.
\item For PMHE2, by optimality $\bm{\Theta}_{2,t}^*\leq \bm{\Xi}_{2,t-N+1/t}(\mathbf{x}_{\Sigma}(t-N+1,\mathbf{x}_0)),\,\,\forall t$.
    Note that the trajectory for~\eqref{eq:collective-update}, stemming from $\hat{\mathbf{x}}_{t-N}=\mathbf{x}_{\Sigma}(t-N,\mathbf{x}_0)$ and with $$\hat{\mathbf{w}}_k=
    \tilde{\mathbf{A}}(\mathbf{x}_{\Sigma}(k,\mathbf{x}_0)-\tilde{\mathbf{x}}_{k/t-1})=
    \tilde{\mathbf{A}}\mathbf{A}^{k-(t-N)}(\mathbf{x}_{\Sigma}(t-N,\mathbf{x}_0)-\hat{\mathbf{x}}_{t-N/t-1})$$
    for $k=t-N,\dots,t-1$ is $\hat{\mathbf{x}}_{k}=\mathbf{x}_{\Sigma}(k,\mathbf{x}_0)$ for $k=t-N+1,\dots,t$, and hence it is a feasible trajectory. Since $\mathbf{y}^{\sharp}_k$, $k=t-N,\dots,t$ corresponds to the deterministic system output from Definition~\ref{def:convergence}, it follows that $$\hat{\mathbf{v}}_k=\mathbf{y}^{\sharp}_k-(\mathbf{C}^*\mathbf{x}_{\Sigma}(k,\mathbf{x}_0)+\tilde{\mathbf{C}}\tilde{\mathbf{x}}_{k/t-1})=
    \tilde{\mathbf{C}}\mathbf{A}^{k-(t-N)}(\mathbf{x}_{\Sigma}(t-N,\mathbf{x}_0)-\hat{\mathbf{x}}_{t-N/t-1})$$
    for all $k=t-N,\dots,t-1$.
    Moreover, by optimality
    $$\begin{array}{l}\bm{\Xi}_{2,t-N+1/t}(\mathbf{x}_{\Sigma}(t-N+1,\mathbf{x}_0))\leq \mathbf{J}_2(t-N,t,{\mathbf{x}}_{\Sigma}(t-N,\mathbf{x}_0),\hat{\mathbf{w}},\hat{\mathbf{v}},\bm{\Gamma}_{2,t-N})\end{array}$$
    From~\eqref{eq:tot_cost_fcn_2}, one has \begin{align}&\mathbf{J}_2(t-N,t,{\mathbf{x}}_{\Sigma}(t-N,\mathbf{x}_0),\hat{\mathbf{w}},\hat{\mathbf{v}},\bm{\Gamma}_{2,t-N})=\nonumber\\
    &=\sum_{k=t-N}^{t-1}\mathbf{L}_2(\hat{\mathbf{w}}_k, \hat{\mathbf{v}}_k)+\bm{\Gamma}_{2,t-N}({\mathbf{x}}_{\Sigma}(t-N,\mathbf{x}_0);{\hat{\mathbf{x}}}_{t-N/t-1})
            \nonumber\end{align}
    In view of~\eqref{eq:tot_step_term} and~\eqref{eq:tot_pen_term} we have that
    \begin{align}\mathbf{J}_2&(t-N,t,{\mathbf{x}}_{\Sigma}(t-N,\mathbf{x}_0),\hat{\mathbf{w}},\hat{\mathbf{v}},\bm{\Gamma}_{2,t-N})=\nonumber\\
    &=\sum_{k=t-N}^{t-1}\frac{1}{2}\|\tilde{\mathbf{C}}\mathbf{A}^{k-(t-N)}(\mathbf{x}_{\Sigma}(t-N,\mathbf{x}_0)-\hat{\mathbf{x}}_{t-N/t-1})\|_{(\mathbf{R}_{k/t-1})^{-1}}^2 +
    \frac{1}{2}\|\tilde{\mathbf{A}}\mathbf{A}^{k-(t-N)}(\mathbf{x}_{\Sigma}(t-N,\mathbf{x}_0)-\hat{\mathbf{x}}_{t-N/t-1})\|_{(\mathbf{Q}_{k/t-1})^{-1}}^2+\nonumber\\
    &+\frac{1}{2}\|\mathbf{x}_{\Sigma}(t-N,\mathbf{x}_0)-{\hat{\mathbf{x}}}_{t-N/t-1}\|^2_{\bm{\Pi}_{t-N/t-1}^{-1}}+\bm{\Theta}^*_{2,t-1}
            \nonumber\\
    &=\frac{1}{2}\|\mathbf{x}_{\Sigma}(t-N,\mathbf{x}_0)-\hat{\mathbf{x}}_{t-N/t-1}\|_{    \sum_{k=t-N}^{t-1}(\mathbf{A}^{k-(t-N)})^T\left(\tilde{\mathbf{C}}^T(\mathbf{R}_{k/t-1})^{-1}\tilde{\mathbf{C}}+\tilde{\mathbf{A}}^T(\mathbf{Q}_{k/t-1})^{-1}\tilde{\mathbf{A}}
    \right)\mathbf{A}^{k-(t-N)}+\bm{\Pi}_{t-N/t-1}^{-1}}^2+\bm{\Theta}^*_{2,t-1}
            \nonumber\end{align}
    and in view of~\eqref{eq:assumption-penalty_PMHE2},
   $\bm{\Theta}_{2,t}^*\leq \bm{\Xi}_{2,t-N/t-1}(\mathbf{x}_{\Sigma}(t-N,\mathbf{x}_0))$. We can iterate this procedure and prove that
   \begin{equation}\bm{\Theta}_{2,t}^*\leq {\Gamma}_{2,0}({\mathbf{x}}_{0})\label{eq:proof_step_2_2}\end{equation}
    for all $t$, for any $\mathbf{x}_{0}\in\mathbb{X}$.
\item For PMHE3, by optimality $\bm{\Theta}_{3,t}^*\leq \bm{\Xi}_{3,t-N+1/t}(\mathbf{x}_{\Sigma}(t-N+1,\mathbf{x}_0)),\,\,\forall t$.
    Note that the trajectory stemming from $\hat{\mathbf{x}}_{t-N}=\mathbf{x}_{\Sigma}(t-N,\mathbf{x}_0)$ with $\hat{\mathbf{w}}_k=0$ (in view of constraints~\eqref{eq:collective constraints_w_for_PMHE3}), for $k>t-N$ is
    $$\begin{array}{lcl}\hat{\mathbf{x}}_k&=&(\mathbf{A}^*)^{k-(t-N)}\mathbf{x}_{\Sigma}(t-N,\mathbf{x}_0) + \sum_{j=0}^{k-(t-N)-1}(\mathbf{A}^*)^{j}\tilde{\mathbf{A}}\tilde{\mathbf{x}}_{k-j-1/t-1}\\
    &=&(\mathbf{A}^*)^{k-(t-N)}\mathbf{x}_{\Sigma}(t-N,\mathbf{x}_0) + \sum_{j=0}^{k-(t-N)-1}(\mathbf{A}^*)^{j}\tilde{\mathbf{A}}\mathbf{A}^{k-j-1-(t-N)}\hat{\mathbf{x}}_{t-N/t-1}
    \end{array}$$
    since $\tilde{\mathbf{x}}_{k/t-1}=\mathbf{A}^{k-(t-N)}\hat{\mathbf{x}}_{t-N/t-1}$. Since $\mathbf{y}^{\sharp}_k$, $k=t-N,\dots,t-1$, corresponds to the deterministic system output from Definition~\ref{def:convergence}, it follows that $$\begin{array}{lcl}\hat{\mathbf{v}}_k&=&\mathbf{y}^{\sharp}_k-(\mathbf{C}^*\hat{\mathbf{x}}_k+\tilde{\mathbf{C}}\tilde{\mathbf{x}}_{k/t-1})\\
    &=&\mathbf{C}^*\left((\mathbf{A}^*)^{k-(t-N)}\mathbf{x}_{\Sigma}(t-N,\mathbf{x}_0) + \sum_{j=0}^{k-(t-N)-1}(\mathbf{A}^*)^{j}\tilde{\mathbf{A}}\mathbf{A}^{k-j-1-(t-N)}\mathbf{x}_{\Sigma}(t-N,\mathbf{x}_0)\right)+\tilde{\mathbf{C}}\mathbf{A}^{k-(t-N)}\mathbf{x}_{\Sigma}(t-N,\mathbf{x}_0)+\\
    &&-\left(\mathbf{C}^*((\mathbf{A}^*)^{k-(t-N)}\mathbf{x}_{\Sigma}(t-N,\mathbf{x}_0) + \sum_{j=0}^{k-(t-N)-1}(\mathbf{A}^*)^{j}\tilde{\mathbf{A}}\mathbf{A}^{k-j-1-(t-N)}\hat{\mathbf{x}}_{t-N/t-1})+\tilde{\mathbf{C}}\mathbf{A}^{k-(t-N)}\hat{\mathbf{x}}_{t-N/t-1}\right)\\
    &=&\left(\mathbf{C}^*\sum_{j=0}^{T_k-1}(\mathbf{A}^*)^{j}\tilde{\mathbf{A}}\mathbf{A}^{T_k-j-1}+\tilde{\mathbf{C}}\mathbf{A}^{T_k}\right)(\mathbf{x}_{\Sigma}(t-N,\mathbf{x}_0)-\hat{\mathbf{x}}_{t-N/t-1})
    \end{array}$$
    where $T_k=k-(t-N)$ for simplicity of notation. It is easy to prove that $\mathbf{C}^*\sum_{j=0}^{T_k-1}(\mathbf{A}^*)^{j}\tilde{\mathbf{A}}\mathbf{A}^{T_k-j-1}+\tilde{\mathbf{C}}\mathbf{A}^{T_k}=\mathbf{C}^{\sharp}
    \mathbf{A}^{T_k}-\mathbf{C}^*(\mathbf{A}^*)^{T_k}$. Therefore
    $$\hat{\mathbf{v}}_k=\left(\mathbf{C}^{\sharp}
    \mathbf{A}^{k-(t-N)}-\mathbf{C}^*(\mathbf{A}^*)^{k-(t-N)}\right)(\mathbf{x}_{\Sigma}(t-N,\mathbf{x}_0)-\hat{\mathbf{x}}_{t-N/t-1})$$
    for all $k=t-N,\dots,t-1$.
    Moreover, by optimality
    $$\begin{array}{l}\bm{\Xi}_{3,t-N+1/t}(\mathbf{x}_{\Sigma}(t-N+1,\mathbf{x}_0))\leq \mathbf{J}_3(t-N,t,{\mathbf{x}}_{\Sigma}(t-N,\mathbf{x}_0),0,\hat{\mathbf{v}},\bm{\Gamma}_{3,t-N})\end{array}$$
    From~\eqref{eq:tot_cost_fcn_2}, one has \begin{align}&\mathbf{J}_3(t-N,t,{\mathbf{x}}_{\Sigma}(t-N,\mathbf{x}_0),0,\hat{\mathbf{v}},\bm{\Gamma}_{3,t-N})=\nonumber\\
    &=\sum_{k=t-N}^{t-1}\mathbf{L}_3(0, \hat{\mathbf{v}}_k)+\bm{\Gamma}_{3,t-N}({\mathbf{x}}_{\Sigma}(t-N,\mathbf{x}_0);{\hat{\mathbf{x}}}_{t-N/t-1})
            \nonumber\end{align}
    In view of~\eqref{eq:tot_step_term3} and~\eqref{eq:tot_pen_term3}
    \begin{align}\mathbf{J}_3&(t-N,t,{\mathbf{x}}_{\Sigma}(t-N,\mathbf{x}_0),0,\hat{\mathbf{v}},\bm{\Gamma}_{3,t-N})=\nonumber\\
    &=\frac{1}{2}\sum_{k=t-N}^{t-1}\|\left(\mathbf{C}^{\sharp}
    \mathbf{A}^{k-(t-N)}-\mathbf{C}^*(\mathbf{A}^*)^{k-(t-N)}\right)(\mathbf{x}_{\Sigma}(t-N,\mathbf{x}_0)-\hat{\mathbf{x}}_{t-N/t-1})\|^2+\frac{\mu}{2}
    \|\mathbf{x}_{\Sigma}(t-N,\mathbf{x}_0)-\hat{\mathbf{x}}_{t-N/t-1}\|^2+\bm{\Theta}^*_{3,t-1}\nonumber\\
    &=\frac{1}{2}\|\left(\mathcal{O}_N-\mathcal{O}_{N}^*\right)(\mathbf{x}_{\Sigma}(t-N,\mathbf{x}_0)-\hat{\mathbf{x}}_{t-N/t-1})\|^2+\frac{\mu}{2}
    \|\mathbf{x}_{\Sigma}(t-N,\mathbf{x}_0)-\hat{\mathbf{x}}_{t-N/t-1}\|^2+\bm{\Theta}^*_{3,t-1}\nonumber\\
    &=\frac{1}{2}\|\mathbf{x}_{\Sigma}(t-N,\mathbf{x}_0)-\hat{\mathbf{x}}_{t-N/t-1}\|^2_{\left(\mathcal{O}_N-\mathcal{O}_{N}^*\right)^T
    \left(\mathcal{O}_N-\mathcal{O}_{N}^*\right)+\mu\,I_n}+\bm{\Theta}^*_{3,t-1}
            \nonumber\end{align}
   and in view of~\eqref{eq:assumption-penalty_PMHE3},
   $\bm{\Theta}_{3,t}^*\leq \bm{\Xi}_{3,t-N/t-1}(\mathbf{x}_{\Sigma}(t-N,\mathbf{x}_0))$. We can iterate this procedure and prove that
   \begin{equation}\bm{\Theta}_{3,t}^*\leq {\Gamma}_{3,0}^*({\mathbf{x}}_{0})\label{eq:proof_step_2_3}\end{equation}
    for all $t$, for any $\mathbf{x}_{0}\in\mathbb{X}$.
\end{itemize}
Finally, from~\eqref{eq:proof_step_1} the sequence $\bm{\Theta}_{r,t}^*$ is increasing for all $r=1,2,3$ and from~\eqref{eq:proof_step_2_1},~\eqref{eq:proof_step_2_2} and~\eqref{eq:proof_step_2_3} it is bounded. Therefore, the sequence $\bm{\Theta}_{r,t}^*$ converges and, from~\eqref{eq:proof_step_1},~\eqref{eq:Lconverges} follows for $r=1,2,3$.
\hfill$\blacksquare$\\
For PMHE1 and PMHE2, in order to prove Lemma~\ref{lemma:step_4}, define the ``unconstrained'' transit cost in a generic sequence $\mathbf{z}_k\in \mathbb{X}$, $k=k_1,\dots,k_2$ as
\begin{equation}
\begin{array}{c}
\bm{\Xi}_{r,[k_1:k_2]/t}^u(\{\mathbf{z}_k\}_{k=k_1}^{k_2})=\min_{\hat{\mathbf{x}}_{t-N},\hat{\mathbf{w}}}
\left\{\mathbf{J}_r(t-N,t,{\hat{\mathbf{x}}}_{t-N},\hat{\mathbf{w}},\hat{\mathbf{v}},\bm{\Gamma}_{r,t-N}) \right.\\\left.\text{ subject to }\eqref{eq:collective-update},~\eqref{eq:collective-out}\text{ and } {\hat{\mathbf{x}}}_{k}
=\mathbf{z}_k\text{ for }k=k_1,\dots,k_2\right\}\end{array}\label{eq:unc-arrival_cost_def}\end{equation}
that, differently from $\bm{\Xi}_{r,[k_1:k_2]/t}$ in~\eqref{eq:arrival_cost_def}, does not account for state constraints. Notice that
\begin{align}\bm{\Xi}_{r,[k_1:k_2]/t}^u(\{\mathbf{z}_k\}_{k=k_1}^{k_2})=\sum_{i=1}^M \Xi_{r,[k_1:k_2]/t}^{[i],u}(\{z^{[i]}_k\}_{k=k_1}^{k_2})\label{eq:sum_transit_costs}\end{align}
where the unconstrained transit cost associated to subsystem $i$ (for PMHE1 and PMHE2) is
\begin{equation}
\Xi_{r,[k_1:k_2]/t}^{[i],u}(\{z^{[i]}_k\}_{k=k_1}^{k_2})=\min_{\hat{x}_{t-N}^{[i]},\hat{w}^{[i]}}\left\{J^{[i]}_r(t-N,t,\hat{x}^{[i]}_{t-N},\hat{w}^{[i]},\hat{v}^{[i]}
,\Gamma^{[i]}_{r,t-N})\right.\\ \left.\text{ subject to~\eqref{eq:system_contraints}  and } {\hat{x}}^{[i]}_{k}
=z^{[i]}_k\text{ for }k=k_1,\dots,k_2\right\}
\label{eq:unc-single-arrival_cost_def}\end{equation}
\begin{remark}\label{remark:KF}
Consider the optimization problem
\begin{equation}
\Theta_{r,t}^{[i],KF}=\min_{\hat{x}_{t-N}^{[i]},\hat{w}^{[i]}}
\left\{J^{[i]}_r(t-N,t,\hat{x}^{[i]}_{t-N},\hat{w}^{[i]},\hat{v}^{[i]}
,\Gamma^{[i]}_{r,t-N})\text{\emph{ subject to~\eqref{eq:system_contraints}  }}\right\}
\label{eq:KF}\end{equation}
and let $\hat{x}^{[i],KF}_{k/t}$, $k=t-N,\dots,t$ be the optimal state
sequence generated by $\hat{x}^{[i],KF}_{t-N/t}$ and the optimal
inputs $\hat{w}^{[i],KF}_{k/t}$. It is well known \cite{RawMayneBook}
that $\hat{x}^{[i],KF}_{k/t}$ are the states produced by the
time-varying Kalman Filter applied to system \eqref{eq:modelS1} with
initial state $\hat{x}^{[i],KF}_{t-N}\sim
\mathcal{N}(\hat{x}^{[i],KF}_{t-N/t-1},\Pi^{[i]}_{t-N/t-1})$ and
assuming uncorrelated noise samples
$w_k^{[i]}\sim\mathcal{N}(0,Q_{k/t-1}^{[i]})$ and
$v_k^{[i]}\sim\mathcal{N}(0,R_{k/t-1}^{[i]})$ and with
variables $\tilde{\mathbf{x}}_{k/t-1}$ playing the role of exogenous
inputs. It is also apparent that $z^{[i]}_k=\hat{x}^{[i],KF}_{k/t}$,
$k=k_1,\dots,k_2$ are the global minimizers of $\Xi^{[i],u}_{r,[k_1,k
_2]/t}(\{z_k^{[i]}\}_{k=k_1}^{k_2})$ and the global minimum is $\Theta_{r,t}^{[i],KF}$.
\end{remark}
The next lemma exploits \eqref{eq:Bound_Rawlings_Rao} for proving a lower bound to the transit cost that will be used in the proof of Lemma~\ref{lemma:step_4}.
\begin{lemma}
\label{lemma:BTC}
For $j=t-N,\dots,t-1$, let
\begin{itemize}
\item $\hat{x}^{[i]}_{t-N/t}, \hat{w}^{[i]}_{j/t}$ be the solution to the MHE-$i$ problem \eqref{eq:min-pb},
\item $\hat{x}^{[i],KF}_{t-N/t}, \hat{w}^{[i],KF}_{j/t}$ be the solution to the MHE-$i$ problem \eqref{eq:KF}.
\end{itemize}
Furthermore, assume that
\begin{equation}
\Xi^{[i],u}_{r,[k_1:k_2]/t}(\{z_k^{[i]}\}_{k=k_1}^{k_2})=
\frac{1}{2}\|Z_{[k_1:k_2]}^{[i]}-\hat{X}_{[k_1:k_2]/t}^{[i],KF}\|_{\mathcal{L}^{-1}}+\Theta_{r,t}^{[i],KF}
\label{eq:unconstr_tr_cost}
\end{equation}
where $Z_{[k_1:k_2]}^{[i]}=(z^{[i]}_{k_1},\dots,z^{[i]}_{k_2})$ and $\hat{X}_{[k_1:k_2]/t}^{[i],KF}=(\hat{x}^{[i],KF}_{k_1/t},\dots,\hat{x}^{[i],KF}_{k_2/t})$
and $\mathcal{L}$ is a suitable positive-definite matrix.
Then
\begin{equation}
\Xi^{[i]}_{r,[k_1:k_2]/t}(\{z_k^{[i]}\}_{k=k_1}^{k_2})\geq
\frac{1}{2}\|Z_{[k_1:k_2]}^{[i]}-\hat{X}_{[k_1:k_2]/t}^{[i]}\|_{\mathcal{L}^{-1}}+\Theta_{r,t}^{[i]}
\label{eq:unconstr/costr_tr_cost}
\end{equation}
where $\hat{X}_{[k_1:k_2]/t}^{[i]}=(\hat{x}^{[i]}_{k_1/t},\dots,\hat{x}^{[i]}_{k_2/t})$.
\end{lemma}
\textbf{Proof of Lemma \ref{lemma:BTC}}\\
The cost $J_r^{[i]}$ is a quadratic function in the variables $\hat{x}^{[i]}_{t-N}$ and $\hat{w}^{[i]}$. Moreover, according to the Remark \ref{remark:KF}, its global minimum under constraints \eqref{eq:modelS1_x} only is given by $\hat{x}^{[i],KF}_{t-N/t}$ and $\hat{w}^{[i],KF}_{k/t}=\hat{x}^{[i],KF}_{k+1/t}-A^{[i]}\hat{x}^{[i],KF}_{k/t}-\tilde{A}^{[i]}\,\tilde{\mathbf{x}}_{k/t-1}$, $k=t-N,\dots,t-1$. for the sake of simplicity, in the following we use the short-hand notation $J_r^{[i]}(\hat{x}^{[i]}_{t-N},\hat{w}^{[i]})$ instead of $J_r^{[i]}(t-N,t,\hat{x}^{[i]}_{t-N},\hat{w}^{[i]},\hat{v}^{[i]}
,\Gamma^{[i]}_{r,t-N})$.\\
Let $\bar{x}^{[i]}_{t-N/t}$ and $\bar{w}^{[i]}_{j/t}$, $j=t-N,\dots,t-1$ be the minimizers of the transit cost $\Xi^{[i]}_{r,[k_1:k_2]/t}(\{z_k^{[i]}\}_{k=k_1}^{k_2})$. Since the MHE-$i$ optimization problem has less constraints than the minimization of the transit cost, and both problems have convex constraints, from Lemma \ref{lemma:Bound} one has
\begin{equation}\begin{array}{lccl}
J_r^{[i]}(\bar{x}^{[i]}_{t-N/t},\bar{w}^{[i]}_{\cdot/t})\geq & \underbrace{J_r^{[i]}(\hat{x}^{[i]}_{t-N/t},\hat{w}^{[i]}_{\cdot/t})}+&
\underbrace{J_r^{[i]}(\bar{x}^{[i]}_{t-N/t}-\hat{x}^{[i]}_{t-N/t}+\hat{x}^{[i],KF}_{t-N/t},
\bar{w}^{[i]}_{\cdot/t}-\hat{w}^{[i]}_{\cdot/t}+\hat{w}^{[i],KF}_{\cdot/t})}
&-\Theta_{r,t}^{[i],KF}\\
&\Theta_{r,t}^{[i],*}&(a)&
\end{array}\label{eq:result_GFT_0}\end{equation}
We now show that $(a)\geq \vartheta_r^{[i]}(\{z_k^{[i]}\}_{k=k_1}^{k_2})$, where
\begin{equation}\label{eq:vartheta_Gianc}
\begin{array}{c}
\vartheta_r^{[i]}(\{z_k^{[i]}\}_{k=k_1}^{k_2})=
\min_{\tilde{x}_{t-N}^{[i]},\tilde{w}^{[i]}}
\left\{J_r^{[i]}(\tilde{x}^{[i]}_{t-N}-\hat{x}^{[i]}_{t-N/t}+\hat{x}^{[i],KF}_{t-N/t},
\tilde{w}^{[i]}_{\cdot}-\hat{w}^{[i]}_{\cdot/t}+\hat{w}^{[i],KF}_{\cdot/t})
\right.\\ \left.\text{ subject to~\eqref{eq:update_x&v}  and } {\tilde{x}}^{[i]}_{k}
=z^{[i]}_k\text{ for }k=k_1,\dots,k_2\right\}
\end{array}\end{equation}
Indeed, the values $\tilde{x}^{[i]}_{k}=\bar{x}^{[i]}_{k/t}$ and $\tilde{w}^{[i]}_{k}=\bar{w}^{[i]}_{k/t}$ are feasible for the optimization problem \eqref{eq:vartheta_Gianc}.\\
We define the following variables for $j=t-N,\dots,t-1$
$$\begin{array}{lcl}
\Delta x_j^{[i]}&=&\tilde{x}^{[i]}_{j}-\hat{x}^{[i]}_{j/t}+\hat{x}^{[i],KF}_{j/t}\\
\Delta w_j^{[i]}&=&\tilde{w}^{[i]}_{j}-\hat{w}^{[i]}_{j/t}+\hat{w}^{[i],KF}_{j/t}\\
\Delta v_j^{[i]}&=&\tilde{v}^{[i]}_{j}-\hat{v}^{[i]}_{j/t}+\hat{v}^{[i],KF}_{j/t}
\end{array}
$$
where $\tilde{x}^{[i]}_{j}$ is the state sequence stemming from $\tilde{x}^{[i]}_{t-N}$ and $\{\tilde{w}^{[i]}_{k}\}_{k=t-N}^{j-1}$ and, from \eqref{eq:modelS1_y}
$$\begin{array}{lcl}
\tilde{v}^{[i]}_{j}&=&y^{[i]}_{j}-{C}^{[i]}\,\tilde{x}^{[i]}_j-\tilde{C}^{[i]}\,\tilde{\mathbf{x}}_{k/t-1}\\
\hat{v}^{[i]}_{j/t}&=&y^{[i]}_{j}-{C}^{[i]}\,\hat{x}^{[i]}_{j/t}-\tilde{C}^{[i]}\,\tilde{\mathbf{x}}_{k/t-1}\\
\hat{v}^{[i],KF}_{j/t}&=&y^{[i]}_{j}-{C}^{[i]}\,\hat{x}^{[i],KF}_{j/t}-\tilde{C}^{[i]}\,\tilde{\mathbf{x}}_{k/t-1}
\end{array}$$
Since system \eqref{eq:modelS1} is linear with respect to inputs $(\hat{w}_{k}^{[i]},\hat{v}_{k}^{[i]},\tilde{\mathbf{x}}_{k/t-1})$, from the superposition principle we have that, for $j=t-N,\dots,t-1$, the variables
$\Delta x_j^{[i]}$, $\Delta w_j^{[i]}$, and $\Delta v_j^{[i]}$ obey to the dynamics \eqref{eq:modelS1}, i.e.,
\begin{equation}
\label{eq:modelS1_Delta}
\begin{array}{lcl}\Delta x_{j+1}^{[i]}&=&A^{[i]}\,\Delta x_j^{[i]}+\tilde{A}^{[i]}\,\tilde{\mathbf{x}}_{j/t-1}+\Delta w_j^{[i]}\\
y^{[i]}_{j}&=&{C}^{[i]}\,\Delta x_j^{[i]}+\tilde{C}^{[i]}\,\tilde{\mathbf{x}}_{j/t-1}+\Delta v_j^{[i]}
\end{array}
\end{equation}
Furthoermore, the constraint $\tilde{x}^{[i]}_k=z_k^{[i]}$, $k=k_1,\dots,k_2$ holds if and only if $\Delta x^{[i]}_k=\Delta z_k^{[i]}$, $k=k_1,\dots,k_2$, where
\begin{equation}\Delta z_k^{[i]}=z^{[i]}_k-\hat{x}^{[i]}_{k/t}+\hat{x}^{[i],KF}_{k/t}\label{eq:vartheta_Gianc_0.5}
\end{equation}
It follows that the optimization problem \eqref{eq:vartheta_Gianc} can be rewritten as
\begin{equation}\label{eq:vartheta_Gianc_3}
\begin{array}{c}
\Delta\vartheta_r^{[i]}(\{\Delta z_k^{[i]}\}_{k=k_1}^{k_2})=
\min_{\Delta x_{t-N}^{[i]},\Delta w^{[i]}}
\left\{J_r^{[i]}(\Delta x^{[i]}_{t-N},\Delta w^{[i]}_{\cdot})
\text{ subject to~\eqref{eq:modelS1_Delta}  and \eqref{eq:vartheta_Gianc_0.5}, }k=k_1,\dots,k_2\right\}
\end{array}\end{equation}
and it holds that $\vartheta_r^{[i]}(\{ z_k^{[i]}\}_{k=k_1}^{k_2})=\Delta\vartheta_r^{[i]}(\{\Delta z_k^{[i]}\}_{k=k_1}^{k_2})$. Note that $\Delta\vartheta_r^{[i]}(\{\Delta z_k^{[i]}\}_{k=k_1}^{k_2})$ is just the unconstrained arrival cost for system \eqref{eq:modelS1_Delta}, that is identical to system \eqref{eq:modelS1}. Therefore one has
$$\Delta\vartheta_r^{[i]}(\{\Delta z_k^{[i]}\}_{k=k_1}^{k_2})=
\Xi^{[i],u}_{r,[k_1,k_2]/t}(\{\Delta z_k^{[i]}\}_{k=k_1}^{k_2})$$
From \eqref{eq:unconstr_tr_cost} one gets
$$\begin{array}{lcl}\Delta\vartheta_r^{[i]}(\{\Delta z_k^{[i]}\}_{k=k_1}^{k_2})
&=&
\frac{1}{2}\|\Delta Z_{[k_1:k_2]}^{[i]}-\hat{X}_{[k_1:k_2]/t}^{[i],KF}\|_{\mathcal{L}^{-1}}+\Theta_{r,t}^{[i],KF}\\
&=&\frac{1}{2}\| Z_{[k_1:k_2]}^{[i]}-\hat{X}_{[k_1:k_2]/t}^{[i]}+\hat{X}_{[k_1:k_2]/t}^{[i],KF}-\hat{X}_{[k_1:k_2]/t}^{[i],KF}\|_{\mathcal{L}^{-1}}+\Theta_{r,t}^{[i],KF}
\end{array}$$
where the last inequality follows from the definition of $\Delta z_k^{[i]}$ in \eqref{eq:vartheta_Gianc_0.5}. From this and from \eqref{eq:result_GFT_0} we infer \eqref{eq:unconstr/costr_tr_cost}.\\\\
\textbf{Proof of Lemma~\ref{lemma:step_4}}\\

\textbf{a) The case of PMHE1}.
First we show that
\begin{align}\Xi^{[i],u}_{1,[t-N+1:t]/t}(\{z^{[i]}_k\}_{k=t-N+1}^t)=\frac{1}{2}\|Z^{[i]}_t-\hat{X}^{[i],KF}_{t}\|^2_{W^{[i]}_t}+\Theta^{[i],KF}_{1,t}
\label{eq:arrival_cost_PMHE1_01}\end{align}
where $Z^{[i]}_t=(z^{[i]}_{t-N+1},\dots,z^{[i]}_{t})$, $\hat{X}^{[i],KF}_{t}=(\hat{x}^{[i],KF}_{t-N+1/t},\dots,\hat{x}^{[i],KF}_{t/t})$, and the matrix $W^{[i]}_t$ can be computed as a function of $\Pi_{k/t-1}^{[i]}$, $Q_{k/t-1}^{[i]}$ and $R_{k/t-1}^{[i]}$, with $k=t-N,\dots,t-1$. Function $J^{[i]}_1$, subject to the constraints ${\hat{x}}^{[i]}_{k}=z^{[i]}_k$ for $k=t-N+1,\dots,t$, is equal to
\begin{align}
\frac{1}{2}\{
\sum_{k=t-N+1}^{t-1}\left(\|z_{k+1}^{[i]}-A^{[i]}z_{k}^{[i]}-
\tilde{A}^{[i]}\hat{\mathbf{x}}_{k/t-1}\|^2_{(Q_{k/t-1}^{[i]})^{-1}}+
\|y_{k}^{[i]}-C^{[i]}z_{k}^{[i]}-\tilde{C}^{[i]}\hat{\mathbf{x}}_{k/t-1}
\|^2_{(R_{k/t-1}^{[i]})^{-1}}\right)+\label{eq:unc_arr_cost_ind_PMHE1_1}
\\
\begin{array}{c}\underbrace{\begin{array}{c}+
\|w^{[i]}_{t-N}\|^2_{(Q_{t-N/t-1}^{[i]})^{-1}}+\\ \vspace{3mm} \\\end{array}
\begin{array}{c}\underbrace{\|y_{t-N}^{[i]}-C^{[i]}\hat{x}_{t-N}^{[i]}
-\tilde{C}^{[i]}\hat{\mathbf{x}}_{t-N/t-1}\|^2_{(R_{t-N/t-1}^{[i]})^{-1}}
+\|\hat{x}^{[i]}_{t-N}-\hat{x}^{[i]}_{t-N/t-1}\|^2_{(\Pi_{t-N/t-1}^{[i]})^{-1}}\}}\\
(a)\end{array}
}\\
(b)\end{array}\nonumber
\end{align}
up to the constant term $\Theta_{1,t-1}^{[i]}$. Considering the term (a), according to Lemma \ref{lemma:opt - smoothing}, it is equivalent to
$$(a)=\|\hat{x}^{[i]}_{t-N}-\bar{x}^{[i]}_{t-N/t-1}\|^2_{(\bar{\Pi}_{t-N/t-1}^{[i]})^{-1}}+\text{const.}$$
where
$$\bar{x}^{[i]}_{t-N/t-1}=\hat{x}^{[i]}_{t-N/t-1}+\mathcal{K}(\Pi^{[i]}_{t-N/t-1},C^{[i]},R^{[i]}_{t-N/t-1})(y^{[i]}_{t-N}-\hat{y}^{[i]}_{t-N/t-1})$$
where $\hat{y}^{[i]}_{t-N/t-1}=C^{[i]}\hat{x}_{t-N/t-1}^{[i]}+\tilde{C}^{[i]}\hat{\mathbf{x}}_{t-N/t-1}$, and $\bar{\Pi}_{t-N/t-1}^{[i]}$ is given by \eqref{eq:unc_arr_cost_ind_PMHE1_2}.
Similarly, according to Lemma \ref{lemma:opt - partial update}, we minimize the partial cost function $(b)$ as follows:
\begin{align}\min(b)&=\min_{\hat{x}_{t-N}^{[i]}}{\|\hat{x}^{[i]}_{t-N}-\bar{x}^{[i]}_{t-N/t-1}\|^2_{(\bar{\Pi}_{t-N/t-1}^{[i]})^{-1}}+\|z_{t-N+1}^{[i]}-A^{[i]}\hat{x}_{t-N}^{[i]}-\tilde{A}^{[i]}\hat{\mathbf{x}}_{t-N/t-1}\|^2_{(Q_{t-N/t-1}^{[i]})^{-1}}+\text{const}}\nonumber\\
&=\|z_{t-N+1}^{[i]}-A^{[i]}\bar{x}_{t-N}^{[i]}-\tilde{A}^{[i]}\hat{\mathbf{x}}_{t-N/t-1}\|^2_{({\Pi}_{t-N/t-1}^{*[i]})^{-1}}+\text{const.}
\label{eq:unc_arr_cost_ind_PMHE1_3}
\end{align}
where ${\Pi}_{t-N/t-1}^{*[i]}$ is given in \eqref{eq:unc_arr_cost_ind_PMHE1_4}.
From~\eqref{eq:unc-single-arrival_cost_def},~\eqref{eq:unc_arr_cost_ind_PMHE1_1} and~\eqref{eq:unc_arr_cost_ind_PMHE1_4} we can write that
\begin{align}
\Xi^{[i],u}_{1,[t-N+1:t]/t}(\{z^{[i]}_k\}_{k=t-N+1}^t)=&\frac{1}{2}\{
\sum_{k=t-N+1}^{t-1}\left(\|z_{k+1}^{[i]}-A^{[i]}z_{k}^{[i]}-\tilde{A}^{[i]}\hat{\mathbf{x}}_{k/t-1}\|^2_{(Q_{k/t-1}^{[i]})^{-1}}+
\|y_{k}^{[i]}-C^{[i]}z_{k}^{[i]}-\tilde{C}^{[i]}\hat{\mathbf{x}}_{k/t-1}\|^2_{(R_{k/t-1}^{[i]})^{-1}}\right)+\nonumber\\
&+\|z_{t-N+1}^{[i]}-A^{[i]}\bar{x}_{t-N}^{[i]}-\tilde{A}^{[i]}\hat{\mathbf{x}}_{t-N/t-1}\|^2_{({\Pi}_{t-N/t-1}^{*[i]})^{-1}}
\}+\Theta_{1,t}^{[i],u}\label{eq:unc_arr_cost_ind_PMHE1_5}
\end{align}
We define the vectors $Y_{1,i}=A^{[i]}\bar{x}^{[i]}_{t-N/t-1}+\tilde{A}^{[i]}\hat{\mathbf{x}}_{t-N/t-1}$,
$$
\begin{array}{lcl}
Y_{2,i}&=&\begin{bmatrix}
\tilde{A}^{[i]}\hat{\mathbf{x}}_{t-N+1/t-1}\\
\vdots\\
\tilde{A}^{[i]}\hat{\mathbf{x}}_{t-1/t-1}
\end{bmatrix}
\end{array},\,
\begin{array}{lcl}
Y_{3,i}&=&\begin{bmatrix}
y_{t-N+1}^{[i]}-\tilde{C}^{[i]}\hat{\mathbf{x}}_{t-N+1/t-1}\\
\vdots\\
y_{t-1}^{[i]}-\tilde{C}^{[i]}\hat{\mathbf{x}}_{t-1/t-1}
\end{bmatrix}
\end{array}
$$
Then, we write~\eqref{eq:unc_arr_cost_ind_PMHE1_5} as follows
\begin{align}
\Xi^{[i],u}_{1,[t-N+1:t]/t}(\{z^{[i]}_k\}_{k=t-N+1}^t)=&\frac{1}{2}\{
\|C_{1,i}Z_{t}^{[i]}-Y_{1,i}\|^2_{({\Pi}_{t-N/t-1}^{*[i]})^{-1}}+
\|C_{2,i}Z_{t}^{[i]}-Y_{2,i}\|^2_{(\mathbf{Q}_{N-1}^{[i]})^{-1}}+
\|Y_{3,i}-C_{3,i}Z_{t}^{[i]}\|^2_{(\mathbf{R}_{N-1}^{[i]})^{-1}}
\}+\text{const.}\label{eq:unc_arr_cost_ind_PMHE1_6}\\
=&\frac{1}{2}
\|C_{4,i}Z_{t}^{[i]}-Y_{4,i}\|^2_{(\tilde{\Pi}_{t-N/t-1}^{[i]})^{-1}}+\text{const.}
\nonumber
\end{align}
where
$Y_{4,i}=(Y_{1,i},Y_{2,i},Y_{3,i})$.
According to Lemma~\ref{lemma:opt - centering}, we have
\begin{equation}
\frac{1}{2}
\|C_{4,i}Z_{t}^{[i]}-Y_{4,i}\|^2_{(\tilde{\Pi}_{t-N/t-1}^{[i]})^{-1}}+\text{const.}=
\frac{1}{2}
\|Z_{t}^{[i]}-\tilde{Y}_{4,i}\|^2_{W_t^{[i]}}+\text{const.}
\label{eq:unc_arr_cost_ind_PMHE1_Gianni}
\end{equation}
where $W_t^{[i]}$ is given by \eqref{eq:Wi_PMHE1_end} and
$$\tilde{Y}_{4,i}=\left(C_{4,i}^T (\tilde{\Pi}_{t-N/t-1}^{[i]})^{-1} C_{4,i}\right)^{-1} C_{4,i}^T (\tilde{\Pi}_{t-N/t-1}^{[i]})^{-1} Y_{4,i}$$
Furthermore, in view of Remark \ref{remark:KF} applied for $k_1=t-N+1$ and $k_2=t$, one has $\hat{X}_{t}^{[i],KF}=\tilde{Y}_{4,i}$ and the constant appearing in \eqref{eq:unc_arr_cost_ind_PMHE1_Gianni} is $\Theta_{1,t}^{[i],KF}$.\\\\
Secondly, we will prove that the left hand side of~\eqref{eq:assumption-penalty_PMHE1} can be written as
$$\sum_{i=1}^{M}\left\{\frac{1}{2}\|Z^{[i]}_{t-1}-\hat{X}^{[i]}_{t-1}\|^{2}_{\mathcal{P}^{[i]}_{t-1}}+\Theta^{[i]}_{1,t-1}\right\}$$
The computation of $\mathcal{P}^{[i]}_{t-1}$ is provided next in two different cases: case I), when $Q^{[i]}_{k/t-1}$ and $R^{[i]}_{k/t-1}$ are computed according to~\eqref{eq:QandR_PMHE1}, and in case II), when $Q^{[i]}_{k/t-1}=Q$ and $R^{[i]}_{k/t-1}=R$.
\begin{itemize}
\item[I)] Assume first that $Q^{[i]}_{k/t-1}$ and $R^{[i]}_{k/t-1}$ ($k=t-N,\dots,t-1$) are computed according to~\eqref{eq:QandR_PMHE1}. Therefore we write
    \begin{subequations}
    \begin{align}
    \mathbf{Q}_{k/t-1}&=\mathbf{Q}+\text{diag}(\tilde{A}^{[1]}\bm{\Delta}_{x,\nu}\bm{\Pi}_{k/t-1}(\tilde{A}^{[1]})^T,\dots,
    \tilde{A}^{[M]}\bm{\Delta}_{x,\nu}\bm{\Pi}_{k/t-1}(\tilde{A}^{[M]})^T)\\
    \mathbf{R}_{k/t-1}&=\mathbf{R}+\text{diag}(\tilde{C}^{[1]}\bm{\Delta}_{y,\nu}\bm{\Pi}_{k/t-1}(\tilde{C}^{[1]})^T,\dots,
    M\tilde{C}^{[M]}\bm{\Delta}_{y,\nu}\bm{\Pi}_{k/t-1}(\tilde{C}^{[M]})^T)
    \end{align}
    \label{eq:QandR_PMHE_condition01}
    \end{subequations}
    Recalling~\eqref{eq:Q_and_R_PMHE1_collective_upperbound} and in view of the Matrix Inversion Lemma, we can write,
    \begin{equation}\begin{array}{lcl}\tilde{\mathbf{A}}^T\mathbf{Q}_{k/t-1}^{-1}\tilde{\mathbf{A}}
    &\leq&\tilde{\mathbf{A}}^T\left(\mathbf{Q}+\tilde{\mathbf{A}}\bm{\Pi}_{k/t-1}\tilde{\mathbf{A}}^T\right)^{-1}\tilde{\mathbf{A}}\\
    &=&\tilde{\mathbf{A}}^T\left(\mathbf{Q}^{-1}-\mathbf{Q}^{-1}\tilde{\mathbf{A}}(\tilde{\mathbf{A}}^T\mathbf{Q}^{-1}\tilde{\mathbf{A}}+\bm{\Pi}_{k/t-1}^{-1})^{-1}\tilde{\mathbf{A}}^T\mathbf{Q}^{-1}\right)\tilde{\mathbf{A}}\\
    &=&\tilde{\mathbf{A}}^T\mathbf{Q}^{-1}\tilde{\mathbf{A}}(\tilde{\mathbf{A}}^T\mathbf{Q}^{-1}\tilde{\mathbf{A}}+\bm{\Pi}_{k/t-1}^{-1})^{-1}\bm{\Pi}_{k/t-1}^{-1}
    \end{array}
    \label{eq:QandR_PMHE_condition02}
    \end{equation}
    Adding and subtracting the term $\bm{\Pi}_{k/t-1}^{-1}(\tilde{\mathbf{A}}^T\mathbf{Q}^{-1}\tilde{\mathbf{A}}+\bm{\Pi}_{k/t-1}^{-1})^{-1}\bm{\Pi}_{k/t-1}^{-1}$ from the right hand side of~\eqref{eq:QandR_PMHE_condition02} we obtain that
    \begin{equation}\begin{array}{lcl}\tilde{\mathbf{A}}^T\mathbf{Q}_{k/t-1}^{-1}\tilde{\mathbf{A}}
    &\leq&(\tilde{\mathbf{A}}^T\mathbf{Q}^{-1}\tilde{\mathbf{A}}+\bm{\Pi}_{k/t-1}^{-1})(\tilde{\mathbf{A}}^T\mathbf{Q}^{-1}\tilde{\mathbf{A}}+\bm{\Pi}_{k/t-1}^{-1})^{-1}\bm{\Pi}_{k/t-1}^{-1}-
    \bm{\Pi}_{k/t-1}^{-1}(\tilde{\mathbf{A}}^T\mathbf{Q}^{-1}\tilde{\mathbf{A}}+\bm{\Pi}_{k/t-1}^{-1})^{-1}\bm{\Pi}_{k/t-1}^{-1}\\
    &\leq&\bm{\Pi}_{k/t-1}^{-1}
    \end{array}
    \label{eq:QandR_PMHE_condition03}
    \end{equation}
    Similarly to~\eqref{eq:QandR_PMHE_condition03}, we can prove that:
        \begin{equation}\begin{array}{lcl}\tilde{\mathbf{C}}^T\mathbf{R}_{k/t-1}^{-1}\tilde{\mathbf{C}}
        &\leq&\bm{\Pi}_{k/t-1}^{-1}
    \end{array}
    \label{eq:QandR_PMHE_condition04}
    \end{equation}
    From~\eqref{eq:QandR_PMHE_condition03} and~\eqref{eq:QandR_PMHE_condition04}, we obtain that the left hand side of~\eqref{eq:assumption-penalty_PMHE1} is
    \begin{align}
    \sum_{k=t-N}^{t-1}&\frac{1}{2}\|\mathbf{z}_k-\hat{\mathbf{x}}_{k/t-1}\|_{
    \tilde{\mathbf{C}}^T(\mathbf{R}_{k/t-1})^{-1}\tilde{\mathbf{C}}+\tilde{\mathbf{A}}^T(\mathbf{Q}_{k/t-1})^{-1}\tilde{\mathbf{A}}
    }^2 +
    \frac{1}{2}\|\mathbf{z}_{t-N}-\hat{\mathbf{x}}_{t-N/t-1}\|^2_{\bm{\Pi}_{t-N/t-1}^{-1}}\nonumber\\
    &\leq\sum_{k=t-N}^{t-1}\frac{1}{2}\|\mathbf{z}_k-\hat{\mathbf{x}}_{k/t-1}\|_{2\bm{\Pi}_{k/t-1}^{-1}}^2 +
    \frac{1}{2}\|\mathbf{z}_{t-N}-\hat{\mathbf{x}}_{t-N/t-1}\|^2_{\bm{\Pi}_{t-N/t-1}^{-1}}\nonumber\\
    &=\sum_{i=1}^{M}\frac{1}{2}\|Z^{[i]}_{t-1}-\hat{X}^{[i]}_{t-1}\|^{2}_{\mathcal{P}^{[i]}_{t-1}}
    \label{eq:QandR_PMHE_condition05}
    \end{align}
    where $\mathcal{P}^{[i]}_{t-1}=($diag$(\frac{1}{3}{\Pi}^{[i]}_{t-N/t-1},\frac{1}{2}{\Pi}^{[i]}_{t-N+1/t-1},\dots,\frac{1}{2}
    {\Pi}^{[i]}_{t-1/t-1}))^{-1}$. From \eqref{eq:arrival_cost_PMHE1_01} and Lemma \ref{lemma:BTC} applied with $k_1=t-N+1$ and $k_2=t$ (and then considering the transit cost obtained at time instant $t-1$ instead of $t$), one has
    $$\Xi^{[i]}_{1,[t-N:t-1]/t-1}(\{z_k^{[i]}\}_{k=t-N}^{t-1})\geq \Theta_{1,t-1}^{[i],*}+\frac{1}{2}\|Z_{t-1}^{[i]}-\hat{X}_{t-1}^{[i]}\|^2_{W_{t-1}^{[i]}}$$
    and if the inequalities \eqref{lmi:PMHE1} hold for $i=1,\dots,M$, in view of \eqref{eq:QandR_PMHE_condition05} also \eqref{eq:assumption-penalty_PMHE1} holds.
\item[II)] Otherwise, we take $\mathbf{Q}_{k/t-1}=\mathbf{Q}$ and that $\mathbf{R}_{k/t-1}=\mathbf{R}$ for all $t$ and $k=t-N,\dots, t-1$. We can write the left hand side of~\eqref{eq:assumption-penalty_PMHE1} as follows
    \begin{align}
    \sum_{i=1}^M&\left\{\sum_{k=t-N}^{t-1}\frac{1}{2}\|\tilde{A}^{[i]}
    (\mathbf{z}_k-\hat{\mathbf{x}}_{k/t-1})\|_{(Q^{[i]})^{-1}}^2 +\frac{1}{2}\|\tilde{C}^{[i]}(\mathbf{z}_k-\hat{\mathbf{x}}_{k/t-1})
    \|_{(R^{[i]})^{-1}}^2+\frac{1}{2}\|z^{[i]}_{t-N}-
    \hat{x}^{[i]}_{t-N/t-1}\|^2_{(\Pi^{[i]}_{t-N/t-1})^{-1}}\right\}
    \nonumber\\
    &=\sum_{i=1}^M\left\{\sum_{k=t-N}^{t-1}\frac{1}{2}\|\sum_{j=1}^M A_{ij}(z^{[j]}_k-\hat{x}^{[j]}_{k/t-1})\|_{(Q^{[i]})^{-1}}^2 +
    \frac{1}{2}\|\sum_{j=1}^M C_{ij}(z^{[j]}_k-\hat{x}^{[j]}_{k/t-1})\|_{(R^{[i]})^{-1}}^2+
   \frac{1}{2}\|z^{[i]}_{t-N}-\hat{x}^{[i]}_{t-N/t-1}\|^2_{(\Pi^{[i]}_{t-N/t-1})^{-1}}\right\}\nonumber\\
    &\leq \sum_{i=1}^M\left\{\sum_{k=t-N}^{t-1}\frac{m_{x}^{[i]}}{2}\sum_{j=1}^M\| A_{ij}(z^{[j]}_k-\hat{x}^{[j]}_{k/t-1})\|_{(Q^{[i]})^{-1}}^2 +
    \frac{m_{y}^{[i]}}{2}\sum_{j=1}^M\| C_{ij}(z^{[j]}_k-\hat{x}^{[j]}_{k/t-1})\|_{(R^{[i]})^{-1}}^2+
    \frac{1}{2}\|z^{[i]}_{t-N}-\hat{x}^{[i]}_{t-N/t-1}\|^2_{(\Pi^{[i]}_{t-N/t-1})^{-1}}\right\}\nonumber\\
    &=\sum_{j=1}^M\left\{\sum_{k=t-N}^{t-1}\frac{1}{2}\| z^{[j]}_k-\hat{x}^{[j]}_{k/t-1}\|_{\sum_{i=1}^M m_x^{[i]} A_{ij}^T(Q^{[i]})^{-1}A_{ij}
    +m_y^{[i]}C_{ij}^T(R^{[i]})^{-1}C_{ij}}^2 +
   \frac{1}{2}\|z^{[j]}_{t-N}-\hat{x}^{[j]}_{t-N/t-1}\|^2_{(\Pi^{[j]}_{t-N/t-1})^{-1}}\right\}\nonumber\\
    &=\sum_{j=1}^{M}\frac{1}{2}\|Z^{[j]}_t-\hat{X}^{[j]}_{t}\|^{2}_{\bar{\mathcal{P}}^{[j]}_{t-1}}
    \label{eq:QandR_PMHE_condition06}
    \end{align}
    Here $A_{ij}$ [resp. $C_{ij}$] is the $j$-th block of $\tilde{A}^{[i]}$ [resp. $\tilde{C}^{[i]}$]. Notice that $A_{ii}=0$ [resp. $C_{ii}=0$] for all $i=1,\dots,M$ and $A_{ij}=0$ [resp. $C_{ij}=0$] if $(j,i)\not\in\mathcal{E}$. Furthermore
    $\bar{\mathcal{P}}^{[j]}_{t-1}=$diag$((\Pi^{[j]}_{t-N/t-1})^{-1}+
    \sum_{i=1}^M (m_x^{[i]}A_{ij}^T(Q^{[i]})^{-1}A_{ij}
    +m_y^{[i]}C_{ij}^T(R^{[i]})^{-1}C_{ij}),\dots,$\break
    $\sum_{i=1}^M (m_x^{[i]}A_{ij}^T(Q^{[i]})^{-1}A_{ij}
    +m_y^{[i]}C_{ij}^T(R^{[i]})^{-1}C_{ij})$. As in the previous case,
    From \eqref{eq:arrival_cost_PMHE1_01} and Lemma \ref{lemma:BTC} applied with $k_1=t-N+1$ and $k_2=t$, one has
    $$\Xi^{[i]}_{1,[t-N:t-1]/t-1}(\{z_k^{[i]}\}_{k=t-N}^{t-1})\geq \Theta_{1,t-1}^{[i],*}+\frac{1}{2}\|Z_{t-1}^{[i]}-\hat{X}_{t-1}^{[i]}\|^2_{(W_{t-1}^{[i]})}$$
    and if the inequalities \eqref{lmi:PMHE1} hold for $i=1,\dots,M$, in view of \eqref{eq:QandR_PMHE_condition06} also \eqref{eq:assumption-penalty_PMHE1} holds.
\end{itemize}

\medskip
\textbf{b) The case of PMHE2}.
First we show that
\begin{align}\Xi^{[i],u}_{2,t-N+1/t}(z^{[i]}_{t-N+1})=\frac{1}{2}\|z^{[i]}_{t-N+1}-\hat{x}^{[i],KF}_{t-N+1}\|^2_{W^{[i]}_t}+\Theta^{[i],KF}_{2,t}
\label{eq:arrival_cost_PMHE2_01}\end{align}
where $W^{[i]}_t$ is given by \eqref{eq:W_PMHE2}.
Function $J^{[i]}_2$, subject to the constraint ${\hat{x}}^{[i]}_{t-N+1}=z^{[i]}_{t-N+1}$, is equal to
\begin{align}
\frac{1}{2}&\{
\sum_{t-N+1}^{t-1}\left(\|\hat{w}_{k}^{[i]}\|^2_{(Q_{k/t-1}^{[i]})^{-1}}+
\|y_{k}^{[i]}-C^{[i]}\hat{x}_{k}^{[i]}-\tilde{C}^{[i]}\tilde{\mathbf{x}}_{k/t-1}\|^2_{(R_{k/t-1}^{[i]})^{-1}}\right)+\label{eq:unc_arr_cost_ind_PMHE2_1}\\
&\begin{array}{c}+\underbrace{\begin{array}{c}\|w^{[i]}_{t-N}\|^2_{(Q_{t-N/t-1}^{[i]})^{-1}}+\end{array}
\|y_{t-N}^{[i]}-C^{[i]}\hat{x}_{t-N}^{[i]}-\tilde{C}^{[i]}\hat{\mathbf{x}}_{t-N/t-1}\|^2_{(R_{t-N/t-1}^{[i]})^{-1}}+\|\hat{x}^{[i]}_{t-N}-\hat{x}^{[i]}_{t-N/t-1}\|^2
_{(\Pi_{t-N/t-1}^{[i]})^{-1}}}\}\\
(b)\end{array}\nonumber
\end{align}
up to the constant term $\Theta_{2,t-1}^{[i]}$. Minimizing the partial cost function $(b)$ we obtain~\eqref{eq:unc_arr_cost_ind_PMHE1_3} and~\eqref{eq:unc_arr_cost_ind_PMHE1_4}.
From~\eqref{eq:unc-single-arrival_cost_def},~\eqref{eq:unc_arr_cost_ind_PMHE1_1} and~\eqref{eq:unc_arr_cost_ind_PMHE1_4} we can write that
\begin{align}
\Xi^{[i],u}_{2,t-N+1/t}&(z^{[i]}_{t-N+1})=\frac{1}{2}\{ \|z_{t-N+1}^{[i]}-A^{[i]}\bar{x}_{t-N}^{[i]}-\tilde{A}^{[i]}\hat{\mathbf{x}}_{t-N/t-1}\|^2_{({\Pi}_{t-N/t-1}^{*[i]})^{-1}}+\nonumber\\
&+\begin{array}{cl}\underbrace{\sum_{t-N+1}^{t-1}\left(\|\hat{w}_{k}^{[i]}\|^2_{(Q_{k/t-1}^{[i]})^{-1}}+
\|y_{k}^{[i]}-C^{[i]}\hat{x}_{k}^{[i]}-\tilde{C}^{[i]}\tilde{\mathbf{x}}_{k/t-1}\|^2_{(R_{k/t-1}^{[i]})^{-1}}\right)}&
\}\\
(c)&
\end{array}
\label{eq:unc_arr_cost_ind_PMHE2_2}
\end{align}
with the constraint $\hat{x}_{t-N+1}^{[i]}={z}_{t-N+1}^{[i]}$, up to a positive constant term. We define $Y_i^{[t-N+1:t-1]}=(y_{t-N+1}^{[i]},\dots,y_{t-1}^{[i]})$, $W_i^{[t-N+1:t-1]}=(w_{t-N+1}^{[i]},\dots,w_{t-1}^{[i]})$, $\mathcal{C}^{[i]}_{w,N-1}$ is given in \eqref{eq:matr-conv_CwN_i},
\begin{align}\mathcal{C}_{N-1}^{[i]}=&\begin{bmatrix}\tilde{C}^{[i]}&0&\dots&0\\
            C^{[i]}\tilde{A}^{[i]}& \tilde{C}^{[i]}&\dots&0\\
            \vdots&\vdots&\ddots&\vdots\\
            C^{[i]}(A^{[i]})^{N-3}\tilde{A}^{[i]}& C^{[i]}(A^{[i]})^{N-4}\tilde{A}^{[i]}&\dots&\tilde{C}^{[i]}
            \end{bmatrix}\label{eq:matr-conv_CN_i}\end{align}
and we write~\eqref{eq:unc_arr_cost_ind_PMHE2_2} as
\begin{align}
\Xi^{[i],u}_{2,t-N+1/t}&(z^{[i]}_{t-N+1})=\frac{1}{2}\{ \|z_{t-N+1}^{[i]}-A^{[i]}\bar{x}_{t-N}^{[i]}-\tilde{A}^{[i]}\hat{\mathbf{x}}_{t-N/t-1}\|^2_{({\Pi}_{t-N/t-1}^{*[i]})^{-1}}+\nonumber\\
&\|W_i^{[t-N+1:t-1]}\|^2_{(\mathbf{Q}^{[i]}_{N-1})^{-1}}+
\|Y_i^{[t-N+1:t-1]}-\mathcal{O}_{N-1}^{[i]} z_{t-N+1}^{[i]}-\mathcal{C}^{[i]}_{N-1}X_i^{[t-N+1:t-1]/t-1}-\mathcal{C}^{[i]}_{w,N-1} W_i^{[t-N+1:t-1]} \|^2_{(\mathbf{R}^{[i]}_{N-1})^{-1}}
\}
\label{eq:unc_arr_cost_ind_PMHE2_3}
\end{align}
up to a positive constant term. If we minimize $(c)$ in~\eqref{eq:unc_arr_cost_ind_PMHE2_3} with respect to $W_i^{[t-N+1:t-1]}$ (see Lemma \ref{lemma:opt - partial update}) we obtain that~\eqref{eq:unc_arr_cost_ind_PMHE2_3} is equal to
\begin{align}
\Xi^{[i],u}_{2,t-N+1/t}&(z^{[i]}_{t-N+1})=\frac{1}{2}\{ \|z_{t-N+1}^{[i]}-A^{[i]}\bar{x}_{t-N}^{[i]}-\tilde{A}^{[i]}\hat{\mathbf{x}}_{t-N/t-1}\|^2_{({\Pi}_{t-N/t-1}^{*[i]})^{-1}}+\nonumber\\
&+
\|Y_i^{[t-N+1:t-1]}-\mathcal{O}_{N-1}^{[i]} z_{t-N+1}^{[i]}-\mathcal{C}^{[i]}_{N-1}X_i^{[t-N+1:t-1]/t-1} \|^2_{({\Pi}_{t-N/t-1}^{**[i]})^{-1}}
\}
\label{eq:unc_arr_cost_ind_PMHE2_4}
\end{align}
where ${\Pi}_{t-N/t-1}^{**[i]}=\mathcal{R}^+(\mathbf{Q}^{[i]}_{N-1},\mathcal{C}^{[i]}_{w,N-1},\mathbf{R}^{[i]}_{N-1})=\mathbf{R}^{[i]}_{N-1}+(\mathcal{C}^{[i]}_{w,N-1})^T\mathbf{Q}^{[i]}_{N-1}\mathcal{C}^{[i]}_{w,N-1}$.\\
In view of Lemma \ref{lemma:opt - smoothing}, we can finally write~\eqref{eq:unc_arr_cost_ind_PMHE2_4} as
\begin{equation}\frac{1}{2}\|z^{[i]}_{t-N+1}-x^{[i],*}\|_{W_t^{[i]}}+\text{const.}
\label{eq:unc_arr_cost_Gian}\end{equation}
where $W^{[i]}_t$ is given by \eqref{eq:W_PMHE2} and
$$x^{[i],*}=A^{[i]}\bar{x}_{t-N}+\tilde{A}^{[i]}\,\tilde{\mathbf{x}}_{k/t-1}+\mathcal{K}({\Pi}_{t-N/t-1}^{*[i]},\mathcal{O}_{N-1}^{[i]},{\Pi}_{t-N/t-1}^{**[i]})
(Y_i^{[t-N+1:t-1]}-A^{[i]}\bar{x}_{t-N}-\tilde{A}^{[i]}\,\tilde{\mathbf{x}}_{k/t-1})
$$
In view of Remark \ref{remark:KF} applied for $k_1=k_2=t-N+1$ one has $\hat{x}^{[i],KF}_{t-N+1/t}=x^{[i],*}$ and the constant appearing in \eqref{eq:unc_arr_cost_ind_PMHE1_Gianni} is $\Theta_{2,t}^{[i],KF}$. This allows
\eqref{eq:unc_arr_cost_ind_PMHE1_Gianni}  to be written as \eqref{eq:arrival_cost_PMHE2_01}.\\\\
Secondly, we will prove that the left hand side of~\eqref{eq:assumption-penalty_PMHE2} can be written as
$$\sum_{i=1}^{M}\left\{\frac{1}{2}\|z^{[i]}_{t-N}-\hat{x}^{[i]}_{t-N/t-1}\|^{2}_{\mathcal{P}^{[i]}_{t-1}}+\Theta^{[i]}_{2,t-1}\right\}$$
The computation of $\mathcal{P}^{[i]}_{t-1}$ is provided next in the cases: I) when $Q^{[i]}_{k/t-1}$ and $R^{[i]}_{k/t-1}$ are computed according to~\eqref{eq:QandR_PMHE2}, and II) when $Q^{[i]}_{k/t-1}=Q$ and $R^{[i]}_{k/t-1}=R$, for $k=t-N,\dots,t-1$.
\begin{itemize}
\item[I)] In this case, $Q^{[i]}_{k/t-1}$ and $R^{[i]}_{k/t-1}$ ($k=t-N,\dots,t-1$) are computed according to~\eqref{eq:QandR_PMHE2}.
    From~\eqref{eq:QandR_PMHE_condition03} and~\eqref{eq:QandR_PMHE_condition04}, we obtain that $\tilde{\mathbf{A}}^T\mathbf{Q}_{k/t-1}^{-1}\tilde{\mathbf{A}}^T\leq \bm{\Pi}^{ol}_{k/t-1}$ and that $\tilde{\mathbf{C}}^T\mathbf{R}_{k/t-1}^{-1}\tilde{\mathbf{C}}^T\leq \bm{\Pi}^{ol}_{k/t-1}$. Therefore the left hand side of~\eqref{eq:assumption-penalty_PMHE2} is
    \begin{align}
    \frac{1}{2}&\|\mathbf{z}_{t-N}-\hat{\mathbf{x}}_{t-N/t-1}\|_{    \sum_{k=t-N}^{t-1}(\mathbf{A}^{k-(t-N)})^T\left(\tilde{\mathbf{C}}^T\mathbf{R}_{k/t-1}^{-1}\tilde{\mathbf{C}}+\tilde{\mathbf{A}}^T\mathbf{Q}_{k/t-1}^{-1}\tilde{\mathbf{A}}
    \right)\mathbf{A}^{k-(t-N)}+\bm{\Pi}_{t-N/t-1}^{-1}}^2\nonumber\\
    &\leq\frac{1}{2}\|\mathbf{z}_{t-N}-\hat{\mathbf{z}}_{t-N/t-1}\|_{
    \sum_{k=t-N}^{t-1}(\mathbf{A}^{k-(t-N)})^T(\bm{\Pi}_{k/t-1}^{ol}+\bm{\Pi}_{k/t-1}^{ol})^{-1}\mathbf{A}^{k-(t-N)} +
    \bm{\Pi}_{t-N/t-1}^{-1}}^2\label{eq:QandR_PMHE2_condition01}
    \end{align}
    Recalling of the expression of $\bm{\Pi}^{ol}_{k/t-1}$ given in~\eqref{eq:QandR_PMHE2_PiOL}, in view of the matrix inversion lemma we can compute that \break
    $(\mathbf{A}^{k-(t-N)})^T(\bm{\Pi}_{k/t-1}^{ol})^{-1}\mathbf{A}^{k-(t-N)}\leq \bm{\Pi}_{t-N/t-1}^{-1}$.
    Therefore
    \begin{align}
    \frac{1}{2}\|\mathbf{z}_{t-N}-\hat{\mathbf{z}}_{t-N/t-1}\|_{
    \sum_{k=t-N}^{t-1}(\mathbf{A}^{k-(t-N)})^T(2\bm{\Pi}_{k/t-1}^{ol})\mathbf{A}^{k-(t-N)} +
    \bm{\Pi}_{t-N/t-1}^{-1}}^2&\leq
    \frac{1}{2}\|\mathbf{z}_{t-N}-\hat{\mathbf{z}}_{t-N/t-1}\|_{
    2\sum_{k=t-N}^{t-1}\bm{\Pi}_{t-N/t-1}^{-1}+
    \bm{\Pi}_{t-N/t-1}^{-1}}^2\nonumber\\
    &= \frac{1}{2}\|\mathbf{z}_{t-N}-\hat{\mathbf{z}}_{t-N/t-1}\|_{
    (2N+1)\bm{\Pi}_{t-N/t-1}^{-1}}^2
    \label{eq:QandR_PMHE2_condition02}
    \end{align}
    Let $\mathcal{P}^{[i]}_{t-1}=(2N+1)(\Pi^{[i]}_{t-N/t-1})^{-1}$. From \eqref{eq:arrival_cost_PMHE2_01} and Lemma \ref{lemma:BTC} applied with $k_1=k_2=t-N+1$  (and then considering the transit cost obtained at time instant $t-1$ instead of $t$), one has
    $$\Xi^{[i]}_{2,t-N/t-1}(z^{[i]}_{t-N})\geq \Theta_{2,t-1}^{[i],*}+\frac{1}{2}\|z_{t-N}^{[i]}-\hat{x}_{t-N}^{[i],KF}\|^2_{W_{t-1}^{[i]}}$$
    and if the inequalities \eqref{lmi:PMHE2} hold for $i=1,\dots,M$, in view of \eqref{eq:QandR_PMHE2_condition02} also \eqref{eq:assumption-penalty_PMHE2} holds.
\item[II)] We set $\mathbf{Q}_{k/t-1}=\mathbf{Q}$ and $\mathbf{R}_{k/t-1}=\mathbf{R}$ for all $t$ and $k=t-N,\dots, t-1$. In this case, the left hand side of~\eqref{eq:assumption-penalty_PMHE2} is
    \begin{align}
    \label{eq:QandR_PMHE2_condition03}
    \frac{1}{2}\|\mathbf{z}_{t-N}-\hat{\mathbf{x}}_{t-N/t-1}\|_{    \sum_{k=t-N}^{t-1}(\mathbf{A}^{k-(t-N)})^T\left(\tilde{\mathbf{C}}^T\mathbf{R}^{-1}\tilde{\mathbf{C}}+\tilde{\mathbf{A}}^T\mathbf{Q}^{-1}\tilde{\mathbf{A}}
    \right)\mathbf{A}^{k-(t-N)}+\bm{\Pi}_{t-N/t-1}^{-1}}^2+\bm{\Theta}^*_{t-1}
    \end{align}
    Our next step is to compute off-line the block-diagonal matrix $\Omega=$diag$(\Omega_1,\dots,\Omega_M)$, where $\Omega_i\in\mathbb{R}^{n_i\times n_i}$ satisfying the inequality
     \begin{align}\sum_{k=t-N}^{t-1}(\mathbf{A}^{k-(t-N)})^T\left(\tilde{\mathbf{C}}^T\mathbf{R}^{-1}\tilde{\mathbf{C}}+\tilde{
     \mathbf{A}}^T\mathbf{Q}^{-1}\tilde{\mathbf{A}}
    \right)\mathbf{A}^{k-(t-N)}\leq \Omega\label{eq:QandR_PMHE2_condition03_LMI}\end{align}
     A simple but conservative choice of $\Omega$ satisfying~\eqref{eq:QandR_PMHE2_condition03_LMI} is
     \begin{align}\Omega=\sum_{k=t-N}^{t-1}\|\mathbf{A}^{k-(t-N)}\|^2_2\|\tilde{\mathbf{C}}^T\mathbf{R}^{-1}\tilde{\mathbf{C}}+\tilde{\mathbf{A}}^T
     \mathbf{Q}^{-1}\tilde{\mathbf{A}}\|_2 I_n
     \label{eq:QandR_PMHE2_condition03_LMI_trivial_choice}\end{align}
     Accordingly, $\Omega_i$ can be set according to \eqref{eq:Omega_i_conservative}.
     From~\eqref{eq:QandR_PMHE2_condition03} and~\eqref{eq:QandR_PMHE2_condition03_LMI} we have that
     \begin{equation}\begin{array}{ll}\frac{1}{2}&\|\mathbf{z}_{t-N}-\hat{\mathbf{x}}_{t-N/t-1}\|_{    \sum_{k=t-N}^{t-1}(\mathbf{A}^{k-(t-N)})^T\left(\tilde{\mathbf{C}}^T\mathbf{R}^{-1}\tilde{\mathbf{C}}+\tilde{\mathbf{A}}^T
     \mathbf{Q}^{-1}\tilde{\mathbf{A}}
    \right)\mathbf{A}^{k-(t-N)}+\bm{\Pi}_{t-N/t-1}^{-1}}^2+\bm{\Theta}^*_{t-1}\\
    &\leq \sum_{i=1}^M\left\{
    \frac{1}{2}\|{z}^{[i]}_{t-N}-\hat{x}^{[i]}_{t-N/t-1}\|_{\Omega_i+(\Pi^{[i]}_{t-N/t-1})^{-1}}^2+{\Theta}^{*[i]}_{t-1}
    \right\}\end{array}\label{eq:QandR_PMHE2_condition02_II}\end{equation}
    Let $\bar{\mathcal{P}^{[i]}}_{t-1}=\Omega_i+(\Pi^{[i]}_{t-N/t-1})^{-1}$. As in the previous case, from \eqref{eq:arrival_cost_PMHE2_01} and Lemma \ref{lemma:BTC} applied with $k_1=k_2=t-N+1$  (and then considering the transit cost obtained at time instant $t-1$ instead of $t$), one has
    $$\Xi^{[i]}_{2,t-N/t-1}(z^{[i]}_{t-N})\geq \Theta_{2,t-1}^{[i],*}+\frac{1}{2}\|z_{t-N}^{[i]}-\hat{x}_{t-N}^{[i],KF}\|^2_{W_{t-1}^{[i]}}$$
    and if the inequalities \eqref{lmi:PMHE2} hold for $i=1,\dots,M$, in view of \eqref{eq:QandR_PMHE2_condition02_II} also \eqref{eq:assumption-penalty_PMHE2} holds.
\end{itemize}\hfill$\blacksquare$

\medskip
\textbf{Proof of Lemma~\ref{lemma:step_4_S3}}\\
If $\mathbf{A}^*$ is invertible, it results that
$$\begin{array}{lcl}\bm{\Xi}_{3,t-N+1/t}(\mathbf{z}_{t-N+1})&=&\mathbf{J}_3(t-N,t,(\mathbf{A}^*)^{-1}(\mathbf{z}_{t-N+1}-\tilde{\mathbf{A}}\hat{\mathbf{x}}_{t-N/t-1}),0,\hat{\mathbf{v}},
\bm{\Gamma}_{3,t-N})\\ &&\text{ subject to system~\eqref{eq:modelS1} with transmission Model~2, where }\hat{\mathbf{w}}_k=0\text{ for }k=t-N,\dots,t-1.\end{array}$$
The term $\bm{\Xi}_{3,t-N+1/t}(\mathbf{z}_{t-N+1})$ can be computed directly.
Collectively, system~\eqref{eq:modelS1}, with transmission Model~2, where $\hat{\mathbf{w}}_k=0$ for $k=t-N,\dots,t-1$ corresponds to
\begin{subequations}
\label{eq:d_const_S3_proof}
\begin{align}
\hat{\mathbf{x}}_{k+1}&=\mathbf{A}^*\hat{\mathbf{x}}_{k}+\tilde{\mathbf{A}}\mathbf{A}^{k-(t-N)}\hat{\mathbf{x}}_{t-N/t-1}\\
\hat{\mathbf{v}}_{k}&=\mathbf{y}^{\sharp}_{k}-\mathbf{C}^*\hat{\mathbf{x}}_{k}-\tilde{\mathbf{C}}\mathbf{A}^{k-(t-N)}\hat{\mathbf{x}}_{t-N/t-1}
\end{align}
\end{subequations}
Therefore, for $k=t-N+1$
$$\hat{\mathbf{v}}_{t-N+1}=\mathbf{y}^{\sharp}_{t-N+1}-\mathbf{C}^*\hat{\mathbf{x}}_{t-N+1}-\tilde{\mathbf{M}}_{1}\hat{\mathbf{x}}_{t-N/t-1}$$
where $\tilde{\mathbf{M}}_{1}=\tilde{\mathbf{C}}\mathbf{A}$. For $k=t-N$
\begin{subequations}
\label{eq:d_const3_S3_proof}
\begin{align}
\hat{\mathbf{x}}_{t-N}&=(\mathbf{A}^*)^{-1}(\hat{\mathbf{x}}^{[i]}_{t-N+1}-\tilde{\mathbf{A}}\hat{\mathbf{x}}_{t-N/t-1})\\
\hat{\mathbf{v}}_{t-N}&=\mathbf{y}^{\sharp}_{t-N}-\mathbf{C}^*(\mathbf{A}^*)^{-1}\hat{\mathbf{x}}_{t-N+1}-\tilde{\mathbf{M}}_{0}\hat{\mathbf{x}}_{t-N/t-1}
\end{align}
\end{subequations}
where $\mathbf{M}_{0}=(\tilde{\mathbf{C}}-\mathbf{C}^*(\mathbf{A}^*)^{-1}\tilde{\mathbf{A}})$.
For $k>t-N+1$
\begin{subequations}
\label{eq:d_const2_S3_proof}
\begin{align}
\hat{\mathbf{x}}_{k}&=(\mathbf{A}^*)^{k-(t-N+1)}\hat{\mathbf{x}}_{t-N+1}+\sum_{j=1}^{k-(t-N+1)}(\mathbf{A}^*)^{j-1}\tilde{\mathbf{A}}\mathbf{A}^{k-j-t+N}\hat{\mathbf{x}}_{t-N/t-1}\\
\hat{\mathbf{v}}_{k}&=\mathbf{y}^{\sharp}_{k}-\mathbf{C}^*(\mathbf{A}^*)^{k-(t-N+1)}\hat{\mathbf{x}}_{t-N+1}-\tilde{\mathbf{M}}_{k-t+N}\hat{\mathbf{x}}_{t-N/t-1}
\end{align}
\end{subequations}
where $\tilde{\mathbf{M}}_{k-t+N}=\mathbf{C}^*\sum_{j=1}^{k-(t-N+1)}(\mathbf{A}^*)^{j-1}\tilde{\mathbf{A}}\mathbf{A}^{k-j-t+N}+\tilde{\mathbf{C}}\mathbf{A}^{k-t+N}$. Then, one obtains
$$\begin{array}{lcl}\bm{\Xi}_{3,t-N+1/t}(\mathbf{z}_{t-N+1})&=&\frac{1}{2}\|\bar{\mathbf{Y}}^{[t-N,t-1]}-\mathcal{M}\hat{\mathbf{x}}_{t-N/t-1}
-\mathcal{O}_{N}^*(\mathbf{A}^*)^{-1}\mathbf{z}_{t-N+1}\|^2+\\
&&+\frac{\mu}{2}\|(\mathbf{A}^*)^{-1}\mathbf{z}_{t-N+1}-\mathbf{P}\hat{\mathbf{x}}_{t-N/t-1}\|^2+\bm{\Theta}_{3,t-1}^{*}\end{array}$$
where $\mathcal{M}=[(\tilde{\mathbf{M}}_0)^T\,\dots\,(\tilde{\mathbf{M}}_{N-1})^T]^T$, $\mathbf{P}=(\mathbf{A}^*)^{-1}\tilde{\mathbf{A}}+I_n$. Therefore, we can write
\begin{align}\bm{\Xi}_{3,t-N+1/t}(\mathbf{z}_{t-N+1})=\frac{1}{2}\|
\mathbf{z}_{t-N+1}-\hat{\mathbf{x}}_{t-N+1/t}\|_{\tilde{\bm{\Pi}}_{t-N+1/t}^{-1}}+\Theta_{3,t}^{*}\label{eq:S3_quadr_form}\end{align}
where $\tilde{\bm{\Pi}}_{t-N+1/t}$ is given by
\begin{equation}\tilde{\bm{\Pi}}_{t-N+1/t}^{-1}=(\mathbf{A}^*)^{-T}\left((\mathcal{O}_{N}^{*})^T \mathcal{O}_{N}^{*}+\mu\,I_{n}\right)(\mathbf{A}^{*})^{-1}\label{eq:Pitilde_S3}\end{equation}
To satisfy~\eqref{eq:assumption-penalty_PMHE3} it is sufficient to chose $\mu$ such that
\begin{align}
\mu\,I_{n}+(\mathcal{O}_N+\mathcal{O}_N^*)^T(\mathcal{O}_N+\mathcal{O}_N^*)\leq \tilde{\bm{\Pi}}_{t-N+1/t}^{-1}
\label{eq:LMI_S3}
\end{align}
Note that $\|\tilde{\bm{\Pi}}_{t-N+1/t}^{-1}\|_2 \geq \sigma^2_{min}((\mathbf{A}^*)^{-1})\sigma_{min}((\mathcal{O}_{N}^*)^T
\mathcal{O}_{N}^*+\mu\,I_{n})$. Recall that $f_{min}=\sigma_{min}(\mathcal{O}_{N}^*)$, $\kappa^{*}=\|\mathbf{A}^*\|_2$ and $\Delta_f=\|\mathcal{O}^*_{N}-\mathcal{O}_{N}\|_2$, and that $\sigma_{min}((\mathbf{A}^{*})^{-1})=\|\mathbf{A}^{*}\|_2^{-1}=1/\kappa^*$. Therefore, $\mu$ has to fulfill
\begin{align}
\mu+\Delta_f^2\leq \frac{\mu+f_{min}^{2}}{\kappa^{*\,2}}
\label{eq:LMI_S3_2}
\end{align}
If $\kappa^*< 1$,~\eqref{eq:LMI_S3_2} is verified if $\mu\geq\max(0,\mu_{min})$, where $\mu_{min}=\frac{\Delta_f^2-f_{min}^2}{1-\kappa^*}$. On the other hand, if ${\kappa}^*\geq 1$,~\eqref{eq:LMI_S3_2} is verified for all $\mu$ satisfying $0\leq \mu\leq \mu_{max}=\frac{f_{min}^2-\Delta_f^2}{\kappa^{*\,2}-1}$, which has a solution only if $f_{min}\geq \Delta_f$.\\
The condition $f_{min}\geq \Delta_f$ can be interpreted as a small coupling condition. \hfill$\blacksquare$

\medskip
\textbf{Proof of Lemma~\ref{lemma:proof_step_3}}\\
First of all, notice that, from the assumptions, for all the PMHE schemes,~\eqref{eq:Lconverges} implies that
\begin{align}\max_{k=t-N,...,t}(\|\hat{\mathbf{v}}_{k/t}\|,\|{\hat{\mathbf{w}}}_{k/t}\|)\stackrel{t\rightarrow\infty}{\longrightarrow } 0\label{eq:max_to-zero}\end{align}
\textbf{Case (a).} In the noiseless case, for any $t$, the output signal is $$\mathbf{y}^{\sharp}_t=\mathbf{C}^{\sharp}\mathbf{x}_{\Sigma}(t,\mathbf{x}_0)=(\mathbf{C}^*+\tilde{\mathbf{C}})\mathbf{x}_{\Sigma}(t,\mathbf{x}_0)$$
        Hence
        \begin{align}
        \hat{\mathbf{v}}_{k/t}&=\mathbf{y}^{\sharp}_t-\mathbf{C}^*{\hat{\mathbf{x}}}_{k/t}-\tilde{\mathbf{C}}{\hat{\mathbf{x}}}_{k/t-1}=
        \mathbf{C}^*\left(\mathbf{x}_{\Sigma}(k,\mathbf{x}_0)-{\hat{\mathbf{x}}}_{k/t}\right)+\tilde{\mathbf{C}}\left(\mathbf{x}_{\Sigma}(k,\mathbf{x}_0)-{\hat{\mathbf{x}}}_{k/t-1}\right)\nonumber\\
        &=\mathbf{C}^*{{\mathbf{\varepsilon}}}_{k/t}+\tilde{\mathbf{C}}{{\mathbf{\varepsilon}}}_{k/t-1}
        \end{align}
        Note that
        \begin{subequations}
            \begin{align}\mathbf{x}_{\Sigma}(k,\mathbf{x}_0)&=
                \mathbf{A}^{* (k-t+N)}\mathbf{x}_{\Sigma}(t-N,\mathbf{x}_0)+\sum_{j=1}^{k-(t-N)}\mathbf{A}^{* (j-1)}\tilde{\mathbf{A}}\mathbf{x}_{\Sigma}(k-j,\mathbf{x}_0),\\
                {\hat{\mathbf{x}}}_{k/t}&=\mathbf{A}^{* (k-t+N)}{\hat{\mathbf{x}}}_{t-N/t}+\sum_{j=1}^{k-(t-N)}\mathbf{A}^{* (j-1)}\tilde{\mathbf{A}}{\hat{\mathbf{x}}}_{k-j/t-1}+\sum_{j=1}^{k-(t-N)}\mathbf{A}^{* (j-1)}\hat{\mathbf{w}}_{k-j/t}.
            \end{align}
            \label{eq:ricurs}
        \end{subequations}
        Similarly to Lemma~3 in~\cite{Rao01}, one obtains
        \small
        \begin{align}\sum_{k=t-N}^{t-1}\|\hat{\mathbf{v}}_{k/t}\|
            &\geq \sum_{k=t-N}^{t-1}\|\mathbf{C}^* \mathbf{A}^{* (k-t+N)}{\mathbf{\varepsilon}}_{t-N/t}+ \mathbf{C}^*\sum_{j=1}^{k-(t-N)}\mathbf{A}^{* (j-1)}\tilde{\mathbf{A}}{\mathbf{\varepsilon}}_{k-j/t-1}+\tilde{\mathbf{C}}{{\mathbf{\varepsilon}}}_{k/t-1}\|-
            \sum_{k=t-N}^{t-1}\|
            \mathbf{C}^*\sum_{j=1}^{k-(t-N)}\mathbf{A}^{* (j-1)}\hat{\mathbf{w}}_{k-j/t}
            \|\label{eq:obs01}\end{align}
            \normalsize
            Note that
            \begin{align}\sum_{k=t-N}^{t-1}\|
            \mathbf{C}^*\sum_{j=1}^{k-(t-N)}\mathbf{A}^{* (j-1)}\hat{\mathbf{w}}_{k-j/t}
            \|\leq \|\mathbf{C}^*\| \sum_{k=t-N}^{t} \sum_{j=1}^{k-(t-N)}\|\mathbf{A}^{* (j-1)}\| \|\hat{\mathbf{w}}_{k-j/t}\|
            \label{eq:bound_2}\end{align}
            From equations~\eqref{eq:obs01} and~\eqref{eq:bound_2} one obtains
            \small\begin{align}\sum_{k=t-N}^{t-1}\|\mathbf{C}^* \mathbf{A}^{* (k-t+N)}{\mathbf{\varepsilon}}_{t-N/t}+ \mathbf{C}^*\sum_{j=1}^{k-(t-N)}\mathbf{A}^{* (j-1)}\tilde{\mathbf{A}}{\mathbf{\varepsilon}}_{k-j/t-1}+\tilde{\mathbf{C}}{{\mathbf{\varepsilon}}}_{k/t-1}\|\leq \sum_{k=t-N}^{t-1}\|\hat{\mathbf{v}}_{k/t}\|+\|\mathbf{C}^*\| \sum_{k=t-N}^{t-1} \sum_{j=1}^{k-(t-N)}\|\mathbf{A}^{* (j-1)}\| \|\hat{\mathbf{w}}_{k-j/t}\|\label{eq:bound_3}\end{align}\normalsize
            Therefore, from~\eqref{eq:max_to-zero}, equation~\eqref{eq:bound_3} implies that
            \small\begin{align}\sum_{k=t-N}^{t-1}\|\mathbf{C}^* \mathbf{A}^{* (k-t+N)}{\mathbf{\varepsilon}}_{t-N/t}+ \mathbf{C}^*\sum_{j=1}^{k-(t-N)}\mathbf{A}^{* (j-1)}\tilde{\mathbf{A}}{\mathbf{\varepsilon}}_{k-j/t-1}+\tilde{\mathbf{C}}{{\mathbf{\varepsilon}}}_{k/t-1}\| =\|\mathcal{O}^*_{N}{\mathbf{\varepsilon}}_{t-N/t}+\mathcal{C}_{N}E_{t-1}\|\stackrel{t\rightarrow \infty}{\longrightarrow}0\label{eq:converge_1}\end{align}\normalsize
            From now on, for simplicity of notation, let $\alpha^j_t$ indicate asymptotically
            vanishing variables, \emph{i.e.}, $\|\alpha^j_t\|\stackrel{t \rightarrow \infty}{\longrightarrow} 0$,
            for all $j\in\mathcal{V}$. We obtain
            \begin{align}
            \mathcal{O}^*_{N}{\mathbf{\varepsilon}}_{t-N/t}+\mathcal{C}_{N}E_{t-1}=\alpha^1_t
            \label{eq:conv_alg01}
            \end{align}
            Since $(\mathbf{A}^*,\mathbf{C}^*)$ is observable by Assumption~\ref{ass:local-observability}, if $N\geq \bar{n}^o$, $\mathcal{O}^*_{N}$ has full column rank $n$.
            Also, from~\eqref{eq:ricurs} we get
            \begin{align}
            E_{t}=M_1{\mathbf{\varepsilon}}_{t-N/t}+M_2 E_{t-1}+\alpha^2_t.
            \label{eq:conv_alg02}
            \end{align}
            From~\eqref{eq:conv_alg01} and~\eqref{eq:conv_alg02}, we obtain the system~\eqref{eq:step05lemma}.

\textbf{Case (b).} The proof of Case (b) is analogous to the proof of Case (a) until equation~\eqref{eq:converge_1}. Then, in case of PMHE2, the term  appearing in~\eqref{eq:converge_1} can be written as follows
            \begin{equation}\sum_{k=t-N}^{t-1}\|\mathbf{C}^* \mathbf{A}^{* (k-t+N)}{\mathbf{\varepsilon}}_{t-N/t}+ \left(\mathbf{C}^*\sum_{j=1}^{k-(t-N)}\mathbf{A}^{* (j-1)}\tilde{\mathbf{A}}\mathbf{A}^{k-j-t+N}+\tilde{\mathbf{C}}\mathbf{A}^{k-t+N}\right){{\mathbf{\varepsilon}}}_{t-N/t-1}\|
            \label{eq:LHS52}\end{equation}
            By direct calculation one can verify that
            \begin{align}\mathbf{M}_{k-t+N}=\mathbf{C}^*\sum_{j=1}^{k-(t-N)}\mathbf{A}^{* (j-1)}\tilde{\mathbf{A}}\mathbf{A}^{k-j-t+N}+\tilde{\mathbf{C}}\mathbf{A}^{k-t+N}=\mathbf{C}^{\sharp}\mathbf{A}^{k-(t-N)}-\mathbf{C}^*(\mathbf{A}^*)^{k-(t-N)}\label{eq:Mk}
            \end{align} Therefore,~\eqref{eq:LHS52} can be written as
            $$\sum_{k=t-N}^{t-1}\|\mathbf{C}^* \mathbf{A}^{* (k-t+N)}{\mathbf{\varepsilon}}_{t-N/t}+ \left(\mathbf{C}^{\sharp}\mathbf{A}^{k-(t-N)}-\mathbf{C}^*(\mathbf{A}^*)^{k-(t-N)}\right){{\mathbf{\varepsilon}}}_{t-N/t-1}\|=
            \|\mathcal{O}^*_{N}{\mathbf{\varepsilon}}_{t-N/t}+ \left(\mathcal{O}_{N}-\mathcal{O}^*_{N}\right){{\mathbf{\varepsilon}}}_{t-N/t-1}\|$$
             Denoting $\alpha_t$ an asymptotically vanishing term such that $\|\alpha_t\|\stackrel{t \rightarrow \infty}{\longrightarrow} 0$, we obtain the equation~\eqref{eq:lemmaS2}.

\textbf{Case (c).} The proof of Case (c) directly follows from the proof of Case (b).\hfill$\blacksquare$

\textbf{Proof of Theorem~\ref{th:main}}\\
First, note that $\mathcal{O}^*_{N}$ has full column rank, in view of Assumption \ref{ass:local-observability}.
To prove (I), pre-multiply both sides of \eqref{eq:step05lemma1} by the pseudo-inverse of $\mathcal{O}^*_{N}$
    $$\begin{array}{c}{\mathbf{\varepsilon}}_{t-N/t}=-\left((\mathcal{O}^*_{N})^T\mathcal{O}^*_{N}\right)^{-1}(\mathcal{O}^*_{N})^T\mathcal{C}_{N}E_{t-1}+\\
    +\left((\mathcal{O}^*_{N})^T\mathcal{O}^*_{N}\right)^{-1}(\mathcal{O}^*_{N})^T\alpha^1_t
    \end{array}$$
    From~\eqref{eq:step05lemma2}, we obtain that $E_{t}=\Phi_1 E_{t-1}+\alpha^3_t$.
    Therefore, convergence of the error to zero is guaranteed if $\Phi_1$ is Schur.

To prove (II) and (III), pre-multiply both sides of \eqref{eq:lemmaS2} by the pseudo-inverse of $\mathcal{O}^*_{N}$
\begin{align}
{\mathbf{\varepsilon}}_{t-N/t}=&\left((\mathcal{O}^*_{N})^T\mathcal{O}^*_{N}\right)^{-1}(\mathcal{O}^*_{N})^T
\left(\mathcal{O}^*_{N}-\mathcal{O}_{N}\right)\times\nonumber\\
&\times\mathbf{A}{\mathbf{\varepsilon}}_{t-N-1/t-1}+
\left((\mathcal{O}^*_{N})^T\mathcal{O}^*_{N}\right)^{-1}(\mathcal{O}^*_{N})^T\alpha_t
\label{eq:lemma2b_pseudoinverse}
\end{align}
Recalling~\eqref{eq:Phi2}, convergence of the error to zero is guaranteed if $\Phi_2$ is Schur.\hfill$\blacksquare$

\textbf{Proof of Theorem~\ref{th:suff_conv_2-3}}\\
Under the assumptions of Lemma~\ref{def:lemmaS2} [resp. Lemma~\ref{def:lemmaS3}] the dynamics of the estimation error of PMHE2 [resp. PMHE3] is given by~\eqref{eq:lemma2b_pseudoinverse}.
Let $e_t=\|{\mathbf{\varepsilon}}_{t-N/t}\|_2$, and $s_t=\|\left((\mathcal{O}^*_{N})^T\mathcal{O}^*_{N}\right)^{-1}(\mathcal{O}^*_{N})^T\alpha_t\|_2$, which satisfies $s_t\rightarrow 0$ as $t\rightarrow\infty$. From \eqref{eq:lemma2b_pseudoinverse} one has
$$\begin{array}{lcl}e_t&\leq& \|\left((\mathcal{O}^*_{N})^T\mathcal{O}^*_{N}\right)^{-1}\|_2\|\mathcal{O}^*_{N}\|_2\times
\\&&\times\|\mathcal{O}^*_{N}-\mathcal{O}_{N}\|_2\|\mathbf{A}\|_2e_{t-1}+s_t\end{array}$$ Recall that $\|\left((\mathcal{O}^*_{N})^T\mathcal{O}^*_{N}\right)^{-1}\|_2=f_{min}^{-2}$. Therefore, if $a_0$ is defined as in \eqref{eq:a_0}, one has that $e_t\rightarrow 0$ as $t\rightarrow \infty$ if $a_0<1$.\hfill$\blacksquare$

\textbf{Proof of Lemma~\ref{def:lemmaS3_unc}}\\
Let $\mathbf{y}^{\sharp}_k=\mathbf{C}^{\sharp}\mathbf{x}_\Sigma(k,\mathbf{x}_0)$. For $r=3$, the cost function~\eqref{eq:tot cost fcn} is
$$\begin{array}{l}\mathbf{J}_3(\cdot)=\frac{1}{2} \sum_{k=t-N}^{t-1}\|\mathbf{C}\mathbf{A}^{k-t+N}\mathbf{x}_\Sigma(t-N,\mathbf{x}_0)+\\
-\mathbf{C}^*(\mathbf{A}^*)^{k-t+N}\hat{\mathbf{x}}_{t-N}-\mathbf{M}_{k-t+N}\hat{\mathbf{x}}_{t-N/t-1}\|^2+\\
+ \frac{\mu}{2} \|\hat{\mathbf{x}}_{t-N}-\hat{\mathbf{x}}_{t-N/t-1}\|^2\end{array}$$
where $\mathbf{M}_{k-t+N}$ is given in~\eqref{eq:Mk}.
Therefore
\begin{equation}
\begin{array}{lcl}\mathbf{J}_3(\cdot)&=&\frac{1}{2} \| \mathcal{O}_{N}\mathbf{x}_\Sigma(t-N,\mathbf{x}_0) +\\
&&- \mathcal{O}^*_{N}\hat{\mathbf{x}}_{t-N} - \left(\mathcal{O}_{N}-\mathcal{O}^*_{N} \right)\hat{\mathbf{x}}_{t-N/t-1} \|^2+\\
&&+ \frac{\mu}{2} \|\hat{\mathbf{x}}_{t-N}-\hat{\mathbf{x}}_{t-N/t-1}\|^2\end{array}
\label{eq:J_3_end}
\end{equation}
From~\eqref{eq:J_3_end}, the unconstrained minimization problem~\eqref{eq:collective_min_pb} is solved by
$$\begin{array}{c}\hat{\mathbf{x}}_{t-N/t}=\left(\mu I_n+(\mathcal{O}^*_{N})^T\mathcal{O}^*_{N}\right)^{-1}\left[
\left(\mu I_n+(\mathcal{O}^*_{N})^T\right.\right.\times\\
\left.\left.\times(\mathcal{O}^*_{N}-\mathcal{O}_{N})\right)\hat{\mathbf{x}}_{t-N/t-1}+
(\mathcal{O}^*_{N})^T\mathcal{O}_{N}{\mathbf{\varepsilon}}_{t-N/t-1}
\right]\end{array}$$
From this, it follows that the dynamics of the estimation error is given by~\eqref{eq:lemmaS3_unc}.\hfill$\blacksquare$

\textbf{Proof of Corollary~\ref{cor:suff_cond_3_unc}}\\
From \eqref{eq:lemmaS3_unc} we obtain
$$\begin{array}{lcl}e_t&\leq& \|\left(\mu\,I_n+(\mathcal{O}^*_{N})^T\mathcal{O}^*_{N}\right)^{-1}\|_2\times\\
&&\times\|\mu\,I_n+(\mathcal{O}^*_{N})^T(\mathcal{O}^*_{N}-\mathcal{O}_{N})\|_2\|\mathbf{A}\|_2e_{t-1}\end{array}$$
Recall that $\|\left(\mu\,I_n+(\mathcal{O}^*_{N})^T\mathcal{O}^*_{N}\right)^{-1}\|_2=(\mu+f_{min}^2)^{-1}$, and that $\|\mu\,I_n+(\mathcal{O}^*_{N})^T(\mathcal{O}^*_{N}-\mathcal{O}_{N})\|_2\leq$\break $\mu+\|\mathcal{O}^*_{N}\|_2\,\|\mathcal{O}^*_{N}-\mathcal{O}_{N}\|_2$
since $\mu\geq 0$. Being $a(\mu)$ defined as in \eqref{eq:amu}, it follows that $e_t\rightarrow 0$ as $t\rightarrow \infty$ if $a(\mu)<1$.\hfill$\blacksquare$
\end{document}